%% file: 0-manuscript.tex
\documentclass[11pt]{article}

\usepackage{epstopdf,float}
\usepackage{xurl}
\usepackage{multirow}
\usepackage{graphics}
\usepackage{amsmath,amsthm}
\usepackage{subfigure}

\usepackage{color, setspace, multirow}
\usepackage[T1]{fontenc}
\usepackage[utf8]{inputenc}
\usepackage{authblk} 
\usepackage{natbib} 
\usepackage{graphicx}
\usepackage{array}
\usepackage{mwe,hyperref}
\usepackage{graphbox, lineno}
\usepackage[export]{adjustbox}
\usepackage[normalem]{ulem}
 \usepackage{enumitem}\usepackage{amssymb}
\DeclareMathAlphabet{\mathbbold}{U}{bbold}{m}{n}

\newcommand{\bfH}{{\bf H}}

\newcommand{\bfX}{{\bf X}}

\newcommand{\bfbeta}{\mbox{\boldmath $\beta$}}

\newcommand{\bfalpha}{\mbox{\boldmath $\alpha$}}

\newcommand{\bfgamma}{\mbox{\boldmath $\gamma$}}

\newcommand{\bfZ}{{\bf Z}}

\usepackage{bm}
\newcommand{\bmy}{{\bm y}}
\usepackage{booktabs}

\newcommand{\fivevdots}{%
  \vbox{\baselineskip=1pt \lineskiplimit=\maxdimen
    \hbox{.}\hbox{.}\hbox{.}\hbox{.}\hbox{.}%
  }%
}

\newcommand\redsout{\bgroup\markoverwith{\textcolor{red}{\rule[0.5ex]{2pt}{0.4pt}}}\ULon}

\newcommand{\blind}{1}

\def\spacingset#1{\renewcommand{\baselinestretch}%
	{#1}\small\normalsize} \spacingset{1}

\date{}

\begin{document}

\if1\blind
{
	\title{Bayesian Nonparametric Approaches to Ordinal Drought Modeling in the United States}
	\author{Mostafa Shams$^{\dagger}$, Robert Erhardt, Staci Hepler\\
		Department of Statistical Sciences, Wake Forest University, U.S.
	}
	\maketitle
} \fi

\if0\blind
{
	\title{\bf A }
	\maketitle
} \fi

\bigskip
\begin{abstract}
Data observed over space and time can exhibit dependence across both dimensions, and datasets can grow very large in both the number of locations and the number of time periods. This dependence and data size mean that many statistical models cannot be fit at a reasonable computational cost as a result of dense matrix inversions, large parameter spaces, or memory and storage challenges.  When the data are ordinal, this only adds to the computational complexity of model fitting.  Many ordinal models rely on a latent continuous variable designed to capture dependence in a computationally efficient manner, and then partitioned according to cutoff parameters to yield the observed ordinal response data. A very common choice is a latent Gaussian distribution, which can accommodate different dependence structures and permits Gibbs sampling in a Bayesian framework. Unfortunately, this model can be overly restrictive and fails to provide the flexibility needed to capture a range of ordinal outcomes. In this paper, we demonstrate the use of Bayesian nonparametric (BNP) methods to enhance the model flexibility of a Gaussian latent model for ordinal drought data observed over space and time, with Dirichlet process priors inducing clustering among time periods within each spatial location. In this work, we model ordinal drought data separately at each spatial location while accounting for temporal dependence, but we do not model spatial dependence across locations. We show that these BNP models often outperform Bayesian parametric approaches at a reasonable computational cost.
  
\end{abstract}

\noindent%
\vfill

$\dagger$ Corresponding author.  Address 1834 Wake Forest Road, Winston-Salem NC, 27109, US.  Email \texttt{shamsm@wfu.edu}

\newpage
\spacingset{1.45} 

\section{Introduction}

\input{1-introduction}

\section{Drought Data}

\input{2-data}

\section{Models}

\input{3-model}

\section{Results}

\input{4-results}

\section{Discussion}

\input{5-discussion}

\section{Data and Code Availability}
This study uses ordinal drought data observed over space and time across the conterminous United States (CONUS), which are publicly available through the Dryad Digital Repository at \url{https://datadryad.org/stash/dataset/doi:10.5061/dryad.g1jwstqw7}. Detailed information on the original data sources, preprocessing steps, and exploratory analyses is provided in \cite{erhardt2024homogenized}. The R code used to implement the models and analyses in this study is publicly available on GitHub at \url{https://github.com/m0stafa-shams/bnp-ordinal-drought}.

\bibliographystyle{jasa}
\bibliography{ref}


\newpage

\clearpage
\appendix
\counterwithin{figure}{section}

\renewcommand{\thetable}{A.\arabic{table}}
\setcounter{table}{0} 

\section{Appendix/Supplemental}

\input{6-appendix}

\end{document}

%% file: 1-introduction.tex
\begin{figure}
\begin{center}
{\includegraphics[width=4.0in]{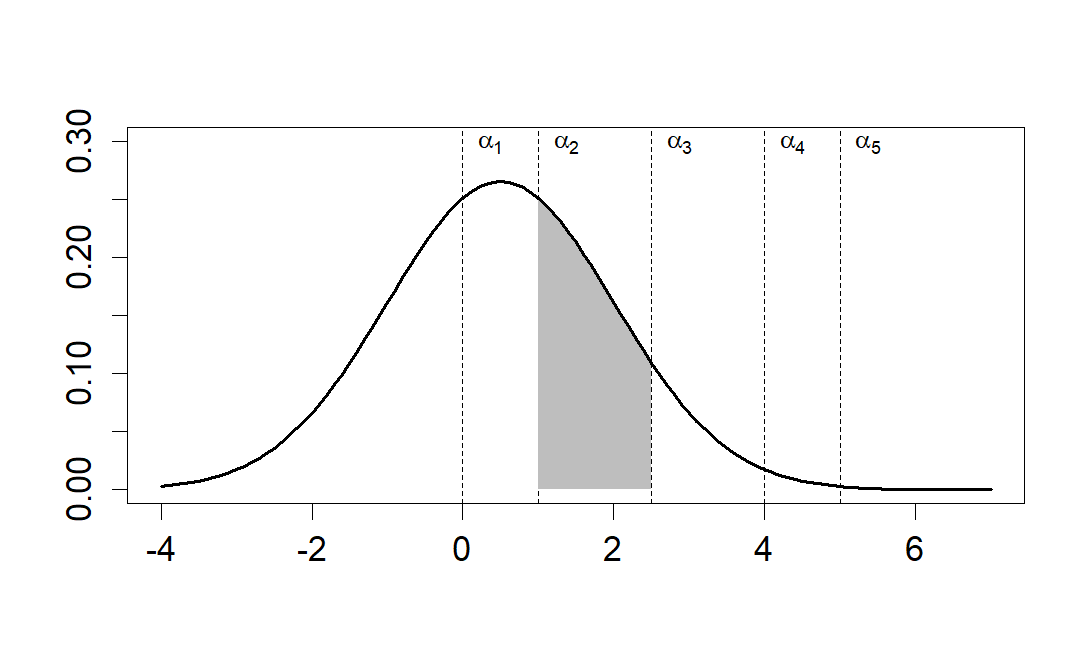}}
{\caption{An underlying latent Gaussian random variable $Z_{i,t}$, when combined with cutoffs $-\infty = \alpha_0 < \alpha_1 < \alpha_2 < \alpha_3 < \alpha_4 < \alpha_5 < \alpha_6 = +\infty$, will give rise to an ordinal variable $Y_{i,t}$ with $J+1=6$ categories.  The shaded area shows $P(Y_{i,t} = 2)$.}\label{fig:latent}}
\end{center}
\end{figure}

Ordinal data observed over space and time are categorical data which take values, $Y_{i,t} \in \{0,1, \dots ,J\}$, where $i$ indexes the $i^{\text{th}}$ spatial location, $t$ indexes $t^{\text{th}}$ time period and the $J+1$ possible categories have a natural ordering. While ordinal data has a discrete support, many models for ordinal data rely on a latent (i.e. unmeasured) continuous process $Z_{i,t} \sim f(z_{i,t})$ \citep{agresti2010analysis, agresti2014some}.  Reasons for preferring a latent variable include parsimony in the parameter space, as it avoids large numbers of parameters to model each categorical level separately, a desire to incorporate dependence structures through the latent variable, a belief in an unobserved continuous construct that gives rise to the observed ordinal category, and many others.  Regardless of the reason, one can recover the ordinal observed variable by partitioning the support of the latent variable with cutoffs $-\infty=\alpha_0 < \alpha_1 < \cdots < \alpha_{J}<\alpha_{J+1}=\infty$, as in
 \begin{equation*}
 Y_{i,t} = \sum_{j=0}^{J} j \cdot I(\alpha_j < Z_{i,t} \leq \alpha_{j+1}),
 \end{equation*}
 where $I(\cdot)$ denotes the indicator function. Gaussian latent variables have a number of appealing properties, including the ability to incorporate different dependence structures through the covariance function \citep{banerjee2014hierarchical}.  A Gaussian latent variable $Z_{i,t} \sim \mathcal{N}(\mu_{i,t}, \tau^2_{i})$ permits the ordinal variable to be defined from the cumulative distribution function of the standard normal $\Phi(\cdot)$. Here, $\tau^2_{i}$ denotes the location-specific precision parameter, defined as the reciprocal of the variance. Setting the precision to be 1 and $\alpha_1 = 0$ for identifiability (there is no loss in generality, as $Z_{i,t}$ is unobserved), this yields 
\[
P(Y_{i,t} = j) = \Phi(\alpha_{j+1} - \mu_{i,t})  - \Phi(\alpha_j - \mu_{i,t}) \hspace{4mm} j=0,1, \dots ,J. \\
\]
An illustrative example is presented in Figure \ref{fig:latent} for the case of $J+1=6$ categories. This model can include covariates to model the mean of $Z_{i,t}$, $\mu_{i,t} = \beta_{0i} + \bfX_{i,t} \bfbeta_{i}$, which changes the center of the latent variable and accordingly redistributes probability across the $J+1$ distinct levels. Here, $\beta_{0i}$ represents the location-specific intercept, and $\bfbeta_{i}$ denotes the location-specific vector of regression coefficients. The precision $\tau^2_{i}$ can be held as a constant, or parameterized with additional covariates as well, to change the spread with a corresponding impact on the ordinal probabilities.  In a Bayesian framework, Gaussian prior distributions for $\beta_{0i}$ and $\bfbeta_{i}$ and a Gamma prior distribution for the precision can permit Gibbs updates.

In principle, the cutoff points $\bfalpha = (-\infty, \alpha_1=0, \alpha_2, \dots, \alpha_J, \infty)$ can be treated as a parameter and estimated from the data; however, in practice this is often very challenging.  Earlier studies have demonstrated that assigning a prior distribution to cutoffs, $\bfalpha$, and modeling this parameter as part of the Bayesian framework can introduce considerable computational complexity \citep{schliep2015data}.  One challenge is the inherent ordering $\alpha_0 < \alpha_1 < \cdots < \alpha_{J} < \alpha_{J+1}$ which sets restrictions in the parameter space, though this can be overcome through transformations.  A larger challenge is that each $\alpha_j$ can only take values in a very narrow range in the Bayesian framework, and accordingly the Markov chain can be very slow to mix.  To illustrate this, consider the conditional distribution of $\alpha_j \mid \bfZ$ for one element $j$.  This must of course be bounded between $\alpha_{j-1}$ and $\alpha_{j+1}$; however, it also must be bounded between the largest $Z_{i,t}$ such that $Y_{i,t} = j$ and the smallest $Z_{i,t}$ such that $Y_{i,t} = j+1$.  To be the valid cutoff separating two adjacent categories, the value of the cutoff would necessarily fall between the largest $Z_{i,t}$ for the level below, and the smallest $Z_{i,t}$ for the level above.  Mathematically, this means that $\alpha_{j} \mid \bfZ$ must be bounded between $\big[ \max(\alpha_{j-1}, \{Z_{i,t} \mid Y_{i,t} = j \} ) , \min(\alpha_{j+1}, \{Z_{i,t} \mid Y_{i,t} = j+1\}) \big]$.  Furthermore, this poor mixing issue only grows as the dimension of the observed data increases in either the number of locations or time periods, as the density of observations grows \citep{erhardt2024spatio}. This can be very limiting, resulting in little variability between sequential draws of a Markov chain regardless of the proposal or prior distributions.   

This challenge can be avoided by treating $\bfalpha$ as fixed and specifying its values in advance, although doing so limits model flexibility in assigning probabilities to the $J+1$ categories. Earlier models that achieve flexibility by treating the cutoff parameters $\bfalpha$ as random typically fix the precision $\tau^2$ for identifiability, whereas our approach instead fixes $\bfalpha$ and allows $\tau^2$ to vary. One can seek model flexibility through the specification of the mean and variance components of the latent variable $Z_{i,t}$.  Examples include \citet{erhardt2024spatio} and \citet{hepler-erhardt2025arxiv}.

\begin{figure}
\begin{center}
{\includegraphics[width=5.0in]{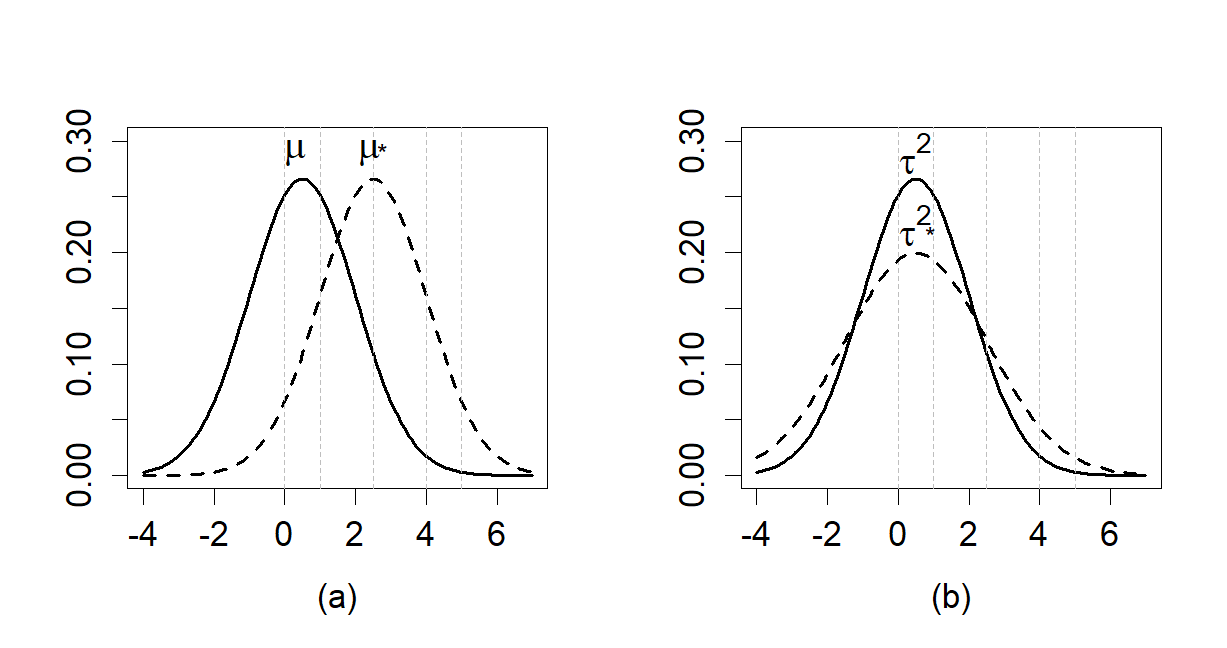}}
{\caption{A demonstration of how a change in the mean $\mu$ or precision $\tau^2$ for a latent Gaussian distribution with fixed cutoff parameters $\bfalpha$ can redistribute probability across $J+1$ categories based on fixed cutoffs $\bfalpha$, but with some restrictions.}\label{fig:latent2}}
\end{center}
\end{figure}

Specifying the mean and variance of the latent variable $Z_{i,t}$ does yield additional model flexibility as shown in Figure \ref{fig:latent2}, but this flexibility is still limited.  Simply stated, shifting the center and/or variance of the Gaussian distribution shown in Figure \ref{fig:latent2} still involves constraints on the relative probabilities that $Y_{i,t}$ can take in each of the $J+1$ ordered levels. For example, when $J+1=6$, there are six category probabilities that must sum to one; however, with fixed cutoffs $\bfalpha$ these probabilities are all determined by only two parameters, the mean and precision. Consequently, they cannot vary independently across categories, which limits the model’s flexibility in capturing more complex ordinal patterns. This is a direct consequence of having both fixed cutoffs $\bfalpha$ and a single Gaussian density for the latent variable.  For a small number of categories of the ordinal variable, this limitation may not be much of an issue.  However, in our application with $J+1 = 6$ categories, we show that the flexibility made possible through specifying the mean and/or variance alone can be insufficient to capture the ordinal distribution we see in the ordinal response $Y_{i,t}$. What is needed is another mechanism to introduce model flexibility.

Advances in Bayesian nonparametric (BNP) methods have introduced flexible modeling frameworks for ordinal data observed over space and time, allowing researchers to move beyond traditional parametric models and better capture complex data structures. For example, \cite{erkanli1993} introduced a Bayesian method for analyzing ordinal data that estimates the link function as a finite mixture of probit links using Dirichlet process mixtures. This approach increases flexibility by allowing the link function to be estimated directly from the data, which is particularly useful when the data indicate a skewed or multimodal density. \cite{kottas2005} and \cite{deYoreo2018} developed Bayesian nonparametric frameworks for multivariate ordinal data and ordinal regression using Dirichlet process mixtures of multivariate normals for the latent responses.

\cite{diLucca2013} introduced a class of Bayesian nonparametric autoregressive models based on dependent Dirichlet process mixtures, providing a flexible framework for modeling either one or multiple time series of observations. They also extended the approach to binary and ordinal data by applying the model to latent variables. \cite{nieto-barajas2014} proposed a Bayesian nonparametric method for clustering time series using a Poisson–Dirichlet process mixture model. Their model accounts for typical features present in a time series like trends, seasonality, and temporal components. 

\cite{mozdzen2022} proposed a Bayesian hierarchical model for spatio-temporal areal data that combines conditional autoregressive priors for spatial and temporal random effects with a Dirichlet process prior to cluster areal units based on location-specific autoregressive and regression parameters, successfully uncovering latent economic patterns in Italian unemployment data. \cite{grazian2024} proposed a novel spatio-temporal stick-breaking process, which defines Dirichlet process weights to incorporate both spatial and temporal dependence. \cite{aiello2025} provides a review of Bayesian nonparametric clustering methods, with a particular focus on their application to spatio-temporal environmental data, such as air pollution. 

In this paper, we describe one way that Bayesian nonparametric methods can achieve additional flexibility, specifically by extending the single latent Gaussian distribution to a mixture of Gaussian distributions, with the number of components to the mixture automatically selected in the methodology as informed by the data. We achieve this by using a Dirichlet process prior on the joint mixing distribution of the time-specific intercept and precision parameters $(\beta_{0,it},\tau^2_{it})$, inducing clustering among time periods within each spatial location. The result of this approach is to permit mixtures of Gaussians which more flexibly adapt to the marginal distributions of ordinal data; in those cases where a single latent Gaussian would suffice, our method chooses a single Gaussian and recovers the simpler model. In this work, we model the ordinal drought data separately at each spatial location and account for temporal dependence through location-specific temporal random effects. We do not model spatial dependence across locations.  

The paper is organized as follows.  Section 2 describes our ordinal data taken from a drought application in the United States, with $J+1=6$ total levels resulting in a need for more model flexibility than can be achieved from a single Gaussian latent distribution for $Z_{i,t}$.  In section 3 we describe our Bayesian nonparametric approach as well as a benchmark Bayesian parametric approach, and derive the log-scoring metric we use to assess model performance.  Section 4 shows extensive results across a number of different model specifications.  This section documents which cases show the nonparametric approach outperforming, and which show essentially a tie in performance.  We conclude with some final remarks in the discussion of section 5.

%% file: 2-data.tex
This study employs ordinal drought data observed over space and time across the conterminous United States (CONUS), which are publicly available at \url{https://datadryad.org/stash/dataset/doi:10.5061/dryad.g1jwstqw7}. For a comprehensive description of the original sources, preprocessing procedures, and exploratory data analysis, refer to \cite{erhardt2024homogenized}. 

The response variable is the US Drought Monitor (USDM), an ordinal indicator of drought severity. The USDM categorizes each region of the United States into one of $J+1 = 6$ ordered levels, ranging from 0 (no drought) to D0 (pre-drought), and then D1 through D4, representing progressively more severe drought conditions. For modeling purposes, we denote the ordinal drought data observed at location $i$ and time period $t$ as $Y_{i,t}$. Each categorical drought level is mapped to an integer in the set $\{0,1,2,3,4, J=5\}$, where 0 $\mapsto 0$, D0 $\mapsto 1$, D1 $\mapsto 2$, D2 $\mapsto 3$, D3 $\mapsto 4$, and D4 $\mapsto 5$. The spatial locations in this study are indexed by $i = 1, \dots , I = 3254$, and the temporal observations are indexed by $t = 1, \dots, T = 522$, corresponding to the unique weeks from 1 January 2011 through 31 December 2020.

This study considers $P = 2$ environmental covariates: weekly average evapotranspiration (\texttt{evp}, [kg/m2]) and a gridded measurement of streamflow over the past 28 days (\texttt{stream28}). The \texttt{evp} covariate was sourced  from the North America Land Data Assimilation System Phase 2 (NLDAS-2), an integrated observation and model reanalysis data set \citep{mitchell2004multi, xia2012continental}. The \texttt{stream28} covariate was separately constructed from in-situ stream gauge data \citep[see,][]{erhardt2024homogenized}. All measurements were obtained on a common spatial grid, consisting of a $0.5^\circ \times 0.5^\circ$ longitude-latitude lattice covering CONUS.

%% file: 3-model.tex
In this section, we present a series of Bayesian hierarchical models developed for analyzing ordinal drought data. We first define the latent variable framework common to all models and then describe the four specific model configurations considered for comparison: two Bayesian parametric (BP) models, which serve as baseline frameworks, and two Bayesian nonparametric (BNP) models, which provide additional flexibility through clustering. The BP models define the baseline structure, while the BNP models extend them by employing Dirichlet process (DP) priors to allow greater flexibility in capturing heterogeneity and distributional complexity at individual spatial locations. Together, these models offer a framework for evaluating and comparing parametric and nonparametric Bayesian approaches to drought modeling. Model predictive performance is subsequently assessed using the full-sample log score criterion to quantify overall predictive accuracy.

\subsection{Latent Variable Framework and Ordinal Mapping}

Let $Y_{i,t} \in \{0,1, \dots ,J\}$ denote the ordinal response variable, which takes one of $J+1$ ordered levels, for location $i=1, \dots , I=3254$ and time period $t=1, \dots, T=522$. We assume there is a latent continuous Gaussian variable $Z_{i,t}$, which is mapped to the observed ordinal response $Y_{i,t}$ via a set of cutoffs:  
$$Y_{i,t} = \sum_{j=0}^J j \cdot \textrm{I}(\alpha_j < Z_{i,t} \leq \alpha_{j+1}),$$ 
where $I(.)$ is the indicator function and $\bfalpha = (-\infty, \alpha_1, \alpha_2, \dots, \alpha_J, \infty)$ represents the set of cutoff points. By definition, $\alpha_0 = -\infty$ and $\alpha_{J+1} = \infty$.

The latent continuous drought variable $Z_{i,t}$ is modeled separately at each spatial location using external environmental covariates. More specifically, we assume:
\begin{equation*} 
Z_{i,t} = \beta_{0i} + \bfX_{i,t} \bfbeta_{i} + \epsilon_{i,t},
\end{equation*}
where $\bfX_{i,t}$ is a $P$-dimensional vector of covariates, $\beta_{0i}$ is a location-specific intercept, $\bfbeta_{i} = (\beta_{1i}, \dots, \beta_{Pi})'$ is a location-specific $P$-dimensional regression coefficients parameter vector, with each $\beta_{pi}$ representing a location-specific covariate effect. The error terms are assumed to be independent and normally distributed as $\epsilon_{i,t} \overset{ind}{\sim}\mathcal{N}(0, \tau^2_{i})$ with location-specific precision parameters $\tau^2_{i}$. 

We set the first cutoff $\alpha_1 = 0$ to ensure model identifiability of the intercept, with no loss in generality \citep[see,][]{feng2014composite, higgs2010clipped, schliep2015data, erhardt2024spatio, hepler-erhardt2025arxiv}. For our application, which includes $J+1 = 6$ ordered drought levels, we fix the entire set of cutoffs as $\bfalpha = (-\infty, \alpha_1=0, \alpha_2=1, \dots, \alpha_J=J-1, \infty)$.

\subsection{Bayesian Parametric Models (Baseline)}

We first consider two Bayesian parametric (BP) models as baseline approaches to analyze drought levels. We specify conjugate priors for the intercept, regression coefficients, and precision parameters, leading to the following Bayesian hierarchical parametric model for the latent variable $Z_{i,t}$. For $i=1,\dots,I$, the model is expressed as:
\begin{equation} \label{eq:bp}
\begin{aligned} 
& Z_{i,t} \mid (\beta_{0i}, \bfbeta_{i}, \tau^2_{i}) \ \overset{ind.}{\sim} \  \mathcal{N}(\beta_{0i} + \bfX_{i,t} \bfbeta_{i}, \tau^2_{i}), \quad \quad t = 1,\dots,T  \\
& \beta_{pi} \ \overset{iid}{\sim} \ \mathcal{N}(\mu_{\beta}, \tau^{2}_{\beta}), \quad \quad p = 0,1,\dots,P  \\
& \tau^2_{i} \ \overset{iid}{\sim} \ \text{Gamma}(a, b).
\end{aligned}  
\end{equation}

In this model, we assign $\mathcal{N}(\mu_{\beta}=0, \tau^{2}_{\beta}=0.04)$ priors to the intercept and regression coefficients $\beta_{pi}$ for $p = 0,1,\dots,P$, and $\text{Gamma}(a=0.01, b=0.01)$ priors to the precision parameters $\tau^2_{i}$. As described in the Drought Data section, our models use $P = 2$ environmental covariates: weekly average evapotranspiration (\texttt{evp}) and a gridded measurement of streamflow over the previous 28 days (\texttt{stream28}). The corresponding regression coefficients for these covariates are denoted by $\beta_1 = \beta_{\texttt{evp}}$ and $\beta_2 = \beta_{\texttt{stream28}}$.

For the second Bayesian parametric model, we account for temporal autocorrelation within each spatial location through a location-specific temporal random effect, while continuing to model the spatial locations separately. The temporal random effect is approximated using $Q$ pairs of Fourier sine and cosine basis functions to achieve dimension reduction and capture temporal dependence. This basis function representation smooths the temporal effect and implicitly induces correlation between nearby time points. The Fourier sine and cosine basis function pairs are particularly useful because they explicitly capture periodic or seasonal patterns present in the drought data. Such patterns are typically driven by seasonal changes in temperature, precipitation, and other environmental variables. Moreover, the Fourier sine and cosine functions form an orthogonal basis that can flexibly approximate a wide range of cyclical behaviors with a relatively small number of parameters, thereby avoiding overfitting and high dimensionality. Specifically, we define:
\begin{equation*} 
Z_{i,t} = \beta_{0i} + \bfX_{i,t} \bfbeta_{i} + \delta_{i,t} + \epsilon_{i,t}, \quad \text{ with } \quad \delta_{i,t} = \bfH_{i,t} \bfgamma_{i},
\end{equation*}
where $\delta_{i,t}$ is the temporal random effect, $\bfH_{i,t}$ is a $2Q$-dimensional vector composed of $Q$ pairs of Fourier sine and cosine basis functions, and $\bfgamma_{i} = \left( \gamma_{1i}^{(c)}, \gamma_{1i}^{(s)}, \dots, \gamma_{Qi}^{(c)}, \gamma_{Qi}^{(s)} \right)'$ is a location-specific $2Q$-dimensional vector of basis coefficients. The linear combination of the Fourier basis functions is expressed as: 
\begin{equation*}
\bfH_{i,t} \bfgamma_{i} = \sum_{q=1}^{Q} \left( \gamma_{qi}^{(c)} \cos\left( \frac{2\pi q t}{R} \right) + \gamma_{qi}^{(s)} \sin\left( \frac{2\pi q t}{R} \right) \right),
\end{equation*}
where $\gamma_{qi}^{(c)}$ and $\gamma_{qi}^{(s)}$ represent the coefficients for the cosine and sine terms, respectively. For weekly drought data, the period $R$ is typically set to the number of weeks in a year, given by $R = 365/7$. Because nearby time points share similar basis function evaluations, this representation induces smooth correlation over time, effectively capturing temporal autocorrelation.

Consequently, we define a Bayesian hierarchical parametric model that accounts for temporal autocorrelation through a temporal random effect represented using Fourier basis functions for the latent variable $Z_{i,t}$. We refer to this model as the BP model with a temporal random effect. For $i=1,\dots,I$, the model is expressed as:
\begin{equation} \label{eq:bp-2q}
\begin{aligned}  
& Z_{i,t} \mid (\beta_{0i}, \bfbeta_{i}, \bfgamma_{i}, \tau^2_{i}) \ \overset{ind.}{\sim} \ \mathcal{N}(\beta_{0i} + \bfX_{i,t} \bfbeta_{i} + \delta_{i,t}, \tau^2_{i}), \quad \quad t = 1,\dots,T  \\ 
& \text{ with } \quad \delta_{i,t} = \bfH_{i,t} \bfgamma_{i}  \\
& \beta_{pi} \ \overset{iid}{\sim} \ \mathcal{N}(\mu_{\beta}, \tau^{2}_{\beta}), \quad \quad p = 0,1,\dots,P  \\
& \gamma_{qi}^{(c)} \ \overset{iid}{\sim} \ \mathcal{N}(\mu_{\gamma}, \tau^{2}_{\gamma}), \quad \quad q = 1,\dots,Q  \\
& \gamma_{qi}^{(s)} \ \overset{iid}{\sim} \ \mathcal{N}(\mu_{\gamma}, \tau^{2}_{\gamma}), \quad \quad q = 1,\dots,Q  \\
& \tau^2_{i} \ \overset{iid}{\sim} \ \operatorname{Gamma}(a, b).
\end{aligned}  
\end{equation}

In this model, we maintain the same prior specifications for the intercept and regression coefficients $\beta_{pi}$ (for $p = 0,1,\dots,P$) and for the precision parameters $\tau^2_{i}$ as described previously. For the basis coefficients associated with the Fourier cosine and sine basis functions, $\gamma_{qi}^{(c)}$ and $\gamma_{qi}^{(s)}$ for $q = 1,\dots,Q$, we assign parametric $\mathcal{N}(\mu_{\gamma}=0, \tau^{2}_{\gamma}=0.04)$ priors.

\subsection{Bayesian Nonparametric Models}

This section presents two Bayesian nonparametric (BNP) models for analyzing drought levels. In particular, we employ the Dirichlet process (DP), originally introduced by \cite{ferguson1973}, which serves as a nonparametric prior for probability distributions. The Dirichlet process is denoted by $\text{DP}(\lambda, G_0)$, where $\lambda > 0$ is a scalar concentration parameter and $G_0$ is the base probability distribution. Due to the stick-breaking construction, as established by \cite{sethuraman1994constructive}, the Dirichlet process yields discrete probability distributions almost surely. To model continuous phenomena, the DP is commonly employed as a prior over the mixing distribution of parameters in a hierarchical model. This formulation gives rise to Dirichlet process mixture models (DPMM) \cite[see][]{neal2000markov}. In our application, the BNP model is fit separately at each spatial location, with the Dirichlet process inducing clustering among time periods within each location. Thus, the clustering is location-specific and does not induce dependence or clustering across spatial locations.

First, we place a DP prior, characterized by the concentration parameter $\lambda_i$ and the base distribution $G_0=G_{0_{\beta_{0}}} \times G_{0_{\tau^2}}$, on the joint mixing distribution of the time-specific intercept and precision parameters $(\beta_{0,it}, \tau^2_{it})$, thereby modeling the latent variable $Z_{i,t}$ using a Dirichlet process mixture model. The remaining parameters are assigned parametric priors, consistent with the specifications in the Bayesian parametric model (\ref{eq:bp}). Furthermore, a Gamma prior is specified for the concentration parameter $\lambda_i$ of the Dirichlet process. This Bayesian nonparametric model, for $i = 1,\dots,I$, is expressed hierarchically as: 
\begin{equation} \label{eq:bnp}
\begin{aligned}  
& Z_{i,t} \mid (\beta_{0,it}, \bfbeta_{i}, \tau^2_{it}) \ \overset{ind.}{\sim} \ \mathcal{N}(\beta_{0,it} + \bfX_{i,t} \bfbeta_{i}, \tau^2_{it}), \quad \quad t = 1,\dots,T  \\
& \beta_{pi} \ \overset{iid}{\sim} \ \mathcal{N}(\mu_{\beta}, \tau^{2}_{\beta}), \quad \quad p = 1,\dots,P  \\
& (\beta_{0,it}, \tau^2_{it}) \mid G_i \ \overset{iid}{\sim} \ G_i, \quad \quad t = 1,\dots,T   \\
& G_i \ \sim \ \text{DP}(\lambda_i, G_0) \\
& G_0 \ = \  G_{0_{\beta_{0}}} \times G_{0_{\tau^2}} \\
& G_{0_{\beta_{0}}} \ = \ \mathcal{N}(\mu_{\beta}, \tau^{2}_{\beta})  \\
& G_{0_{\tau^2}} \ = \ \text{Gamma}(a, b)  \\
& \lambda_i \ \sim \ \text{Gamma}(a_{\lambda},b_{\lambda}). 
\end{aligned}  
\end{equation}

Specifically, $G_{0_{\beta_{0}}} \ = \ \mathcal{N}(\mu_{\beta}=0, \tau^{2}_{\beta}=0.04)$ is defined as the base distribution for the intercept, and $G_{0_{\tau^2}} \ = \ \text{Gamma}(a=0.01, b=0.01)$ as the base distribution for the precision. Note that the index $p$ for $\beta_{pi}$ begins at $1$ because the time-specific intercept and precision parameters $(\beta_{0,it},\tau^2_{it})$ are jointly assigned a DP prior, while the regression coefficients $\beta_{pi}$, for $p=1,\dots,P$, are assigned parametric $\mathcal{N}(\mu_{\beta}=0,\tau^2_{\beta}=0.04)$ priors.

The concentration parameter $\lambda_i$ is assigned a Gamma prior with a shape parameter $a_{\lambda}=1$ and a rate parameter $b_{\lambda}=1$. This parameter controls the clustering behavior of the DP; a larger $\lambda_i$ favors a greater number of  clusters, while a smaller $\lambda_i$ results in fewer clusters. This $\text{Gamma}(1,1)$ prior on $\lambda_i$ is considered weakly informative and is a common choice in the BNP literature, tending to favor models with fewer clusters. However, due to its mean of $1$, this prior does not strongly bias $\lambda_i$ towards either extremely large (many clusters) or extremely small (very few clusters) values. Thus, it is particularly suitable for scenarios lacking strong prior information about the expected number of clusters. Consequently, it is often adopted as a default prior for the concentration parameter in Dirichlet process mixture models.

For the second Bayesian nonparametric model, we assign a DP prior to the joint mixing distribution of the time-specific intercept and precision parameters $(\beta_{0,it}, \tau^2_{it})$ in model (\ref{eq:bp-2q}), which includes a temporal random effect represented using $Q$ pairs of Fourier sine and cosine basis functions. As in the corresponding BP model, each spatial location is modeled separately, with temporal dependence within each location captured through the location-specific temporal random effect. We refer to this model as the BNP model with a temporal random effect. Accordingly, the latent variable $Z_{i,t}$ is modeled using a Dirichlet process mixture model. For $i=1,\dots,I$, the model is expressed as:
\begin{equation} \label{eq:bnp-2q}
\begin{aligned}  
& Z_{i,t} \mid (\beta_{0,it}, \bfbeta_{i}, \bfgamma_{i}, \tau^2_{it}) \ \overset{ind.}{\sim} \ \mathcal{N}(\beta_{0,it} + \bfX_{i,t} \bfbeta_{i} + \delta_{i,t}, \tau^2_{it}), \quad \quad t = 1,\dots,T \\
& \text{ with } \quad \delta_{i,t} = \bfH_{i,t} \bfgamma_{i}  \\
& \beta_{pi} \ \overset{iid}{\sim} \ \mathcal{N}(\mu_{\beta}, \tau^{2}_{\beta}), \quad \quad p = 1,\dots,P  \\
& \gamma_{qi}^{(c)} \ \overset{iid}{\sim} \ \mathcal{N}(\mu_{\gamma}, \tau^{2}_{\gamma}), \quad \quad q = 1,\dots,Q  \\
& \gamma_{qi}^{(s)} \ \overset{iid}{\sim} \ \mathcal{N}(\mu_{\gamma}, \tau^{2}_{\gamma}), \quad \quad q = 1,\dots,Q  \\
& (\beta_{0,it}, \tau^2_{it}) \mid G_i \ \overset{iid}{\sim} \ G_i, \quad \quad t = 1,\dots,T   \\
& G_i \ \sim \ \text{DP}(\lambda_i, G_0) \\
& G_0 \ = \  G_{0_{\beta_{0}}} \times G_{0_{\tau^2}} \\
& G_{0_{\beta_{0}}} \ = \ \mathcal{N}(\mu_{\beta}, \tau^{2}_{\beta})  \\
& G_{0_{\tau^2}} \ = \ \text{Gamma}(a, b)  \\
& \lambda_i \ \sim \ \text{Gamma}(a_{\lambda},b_{\lambda}). 
\end{aligned}  
\end{equation}

As previously stated, the Dirichlet process generates discrete probability distributions almost surely. Consequently, the location-specific mixing distribution $G_i$ in the BNP models (\ref{eq:bnp}) and (\ref{eq:bnp-2q}) is discrete, which implies a positive probability of ties among the time-specific intercept and precision parameters assigned DP priors. These parameter ties naturally give rise to a clustering structure within each spatial location, characterized by $K_i$ distinct parameter pairs, thereby inducing a partition of the time periods into clusters.

\subsection{Predictive Performance Comparison of BNP and BP Models}
To evaluate and compare the predictive performance of the BNP and BP models, we employ the full-sample \textit{log score} criterion, which assesses a model's predictive performance based on the logarithm of the heights of the posterior predictive distributions evaluated at the observed data points. Higher log score values indicate superior predictive performance. Given a dataset $\bmy = (y_1, y_2, \dots, y_T)$ consisting of $T$ observations, and for a specific model $\mathcal{M}$, the full-sample log score \citep[see, e.g.,][]{krnjajic2008,laud1995} is defined as: 
\begin{equation} \label{eq:logscore}
    \text{LS}(\mathcal{M} \mid \bmy) = \frac{1}{T} \sum_{j=1}^T \log \Big(p(y_j \mid \bmy, \mathcal{M})\Big),
\end{equation}
where $p(\cdot)$ denotes the posterior predictive distribution function.

Here, we demonstrate the computation of the full-sample log score for the BP and BNP models with a temporal random effect represented using basis functions. The log scores for models without a temporal random effect are obtained similarly. Let $\bmy_i = (y_{i,1}, y_{i,2}, \dots, y_{i,T})$ denote the observed ordinal drought levels for location $i$, where each observation $y_{i,t} \in \{0,1, \dots ,J\}$ corresponds to one of $J + 1$ ordered drought categories observed at time $t$. The set of cutoffs is fixed as $\bfalpha = (-\infty, \alpha_1=0, \alpha_2=1, \dots, \alpha_J=J-1, \infty)$. For the BP model with a temporal random effect, $\mathcal{M}_{\text{BP}}$, specified in equation (\ref{eq:bp-2q}), the log score for location $i$ is computed as: 

{\footnotesize
\begin{equation} \label{eq:log-calc}
\begin{aligned}
& \text{LS}(\mathcal{M}_{\text{BP}} \mid \bmy_i) = \frac{1}{T} \sum_{t=1}^T \log \Big(p(y_{i,t} \mid \bmy_i, \mathcal{M_{\text{BP}}})\Big) \\
& = \frac{1}{T} \sum_{t=1}^T \log \Bigg(
\Pr(Z_{i,t} < 0)^{I(y_{i,t}=0)} \times \Pr(0 < Z_{i,t} < 1)^{I(y_{i,t}=1)}  \\
& \hspace{2cm} \times \Pr(1 < Z_{i,t} < 2)^{I(y_{i,t}=2)} \times \dots \times \Pr(J-1 < Z_{i,t})^{I(y_{i,t}=J)} \Bigg) \\
& = \frac{1}{T} \sum_{t=1}^T \Bigg( 
I(y_{i,t}=0) \times \log (\Pr(Z_{i,t} < 0)) + I(y_{i,t}=1) \times \log (\Pr(0 < Z_{i,t} < 1)) \\
& \hspace{1cm}  + I(y_{i,t}=2) \times \log (\Pr(1 < Z_{i,t} < 2)) + \dots + I(y_{i,t}=J) \times \log (\Pr(J-1 < Z_{i,t})) \Bigg).
\end{aligned}
\end{equation}
}

\noindent Each probability term $\Pr(\cdot)$ is approximated by averaging over the posterior samples obtained from the Markov chain Monte Carlo (MCMC) sampling algorithm. Let $L$ be the number of MCMC iterations after burn-in and thinning. Denote by $\beta_{0i}^{(l)}$, $\bfbeta_{i}^{(l)}$, $\bfgamma_{i}^{(l)}$, and $\tau^{2^{(l)}}_{i}$ the posterior samples at the $l^{\text{th}}$ iteration, for $l = 1, \dots, L$. The log score for the BP model, $\mathcal{M}_{\text{BP}}$, at location $i$ is then approximated as:

{\footnotesize
\begin{equation} \label{eq:log-calc-bp}
\begin{aligned}
& \text{LS}(\mathcal{M}_{\text{BP}} \mid \bmy_i) \approx \frac{1}{T} \sum_{t=1}^T \left(
I(y_{i,t}=0) \times \log \left(\frac{1}{L} \sum_{l=1}^L \Phi \left(\frac{0- (\beta_{0i}^{(l)} + \bfX_{i,t} \bfbeta_{i}^{(l)} + \bfH_{i,t} \bfgamma_{i}^{(l)})}{1/\sqrt{\tau^{2^{(l)}}_{i}}}\right) \right) \right. \\
& \left. + I(y_{i,t}=1) \times \log \left(\frac{1}{L} \sum_{l=1}^L \left( \Phi \left(\frac{1 - (\beta_{0i}^{(l)} + \bfX_{i,t} \bfbeta_{i}^{(l)} + \bfH_{i,t} \bfgamma_{i}^{(l)})}{1/\sqrt{\tau^{2^{(l)}}_{i}}}\right) - \Phi \left(\frac{0- (\beta_{0i}^{(l)} + \bfX_{i,t} \bfbeta_{i}^{(l)} + \bfH_{i,t} \bfgamma_{i}^{(l)})}{1/\sqrt{\tau^{2^{(l)}}_{i}}}\right) \right) \right) \right. \\
& \left. + I(y_{i,t}=2) \times \log \left(\frac{1}{L} \sum_{l=1}^L \left( \Phi \left(\frac{2 - (\beta_{0i}^{(l)} + \bfX_{i,t} \bfbeta_{i}^{(l)} + \bfH_{i,t} \bfgamma_{i}^{(l)})}{1/\sqrt{\tau^{2^{(l)}}_{i}}}\right) - \Phi \left(\frac{1 - (\beta_{0i}^{(l)} + \bfX_{i,t} \bfbeta_{i}^{(l)} + \bfH_{i,t} \bfgamma_{i}^{(l)})}{1/\sqrt{\tau^{2^{(l)}}_{i}}}\right) \right) \right) \right. \\[3mm]
& \hspace{7cm} \fivevdots \\[3mm]
& \left. + I(y_{i,t}=J) \times \log \left(\frac{1}{L} \sum_{l=1}^L \left( 1 - \Phi \left(\frac{J-1 - (\beta_{0i}^{(l)} + \bfX_{i,t} \bfbeta_{i}^{(l)} + \bfH_{i,t} \bfgamma_{i}^{(l)})}{1/\sqrt{\tau^{2^{(l)}}_{i}}}\right) \right) \right) \right),
\end{aligned}
\end{equation}
}
where $\Phi(\cdot)$ denotes the cumulative distribution function of the standard normal distribution, and $I(\cdot)$ is the indicator function. 

The computation of the log score for the BNP model with a temporal random effect specified in equation (\ref{eq:bnp-2q}), $\mathcal{M}_{\text{BNP}}$, is similar to that of the BP model. A key difference is that the latent variable $Z_{i,t}$ in the BNP model follows a mixture of normal distributions, with the mixture structure varying across MCMC iterations. Notably, the time-specific intercept and precision parameters $(\beta_{0,it}, \tau^2_{it})$ are jointly assigned a DP prior, while the remaining parameters are modeled using parametric priors. Let $L$ denote the number of MCMC iterations after burn-in and thinning, and let $K_i^{(l)}$ represent the number of distinct mixture components in the $l^{\text{th}}$ MCMC iteration. Denote by $\beta_{0i}^{(l,k)}$ and $\tau^{2^{(l,k)}}_{i}$ the posterior samples for the $k^{\text{th}}$ mixture component at iteration $l$, and by $\bfbeta_{i}^{(l)}$ and $\bfgamma_{i}^{(l)}$ the parametric posterior samples at iteration $l$. Then, the full-sample log score is approximated as:

{\footnotesize
\begin{equation} \label{eq:log-calc-bnp}
\begin{aligned}
& \text{LS}(\mathcal{M}_{\text{BNP}} \mid \bmy_i) \approx \frac{1}{T} \sum_{t=1}^T \left(
I(y_{i,t}=0) \times \log \left(\frac{1}{L} \sum_{l=1}^L \sum_{k=1}^{K_i^{(l)}} w_{ik}^{(l)} \Phi \left(\frac{0- (\beta_{0i}^{(l,k)} + \bfX_{i,t} \bfbeta_{i}^{(l)} + \bfH_{i,t} \bfgamma_{i}^{(l)})}{1/\sqrt{\tau^{2^{(l,k)}}_{i}}}\right) \right) \right. \\
& \left. + I(y_{i,t}=1) \times \log \left(\frac{1}{L} \sum_{l=1}^L \sum_{k=1}^{K_i^{(l)}} w_{ik}^{(l)} \left( \Phi \left(\frac{1 - (\beta_{0i}^{(l,k)} + \bfX_{i,t} \bfbeta_{i}^{(l)} + \bfH_{i,t} \bfgamma_{i}^{(l)})}{1/\sqrt{\tau^{2^{(l,k)}}_{i}}}\right) - \Phi \left(\frac{0- (\beta_{0i}^{(l,k)} + \bfX_{i,t} \bfbeta_{i}^{(l)} + \bfH_{i,t} \bfgamma_{i}^{(l)})}{1/\sqrt{\tau^{2^{(l,k)}}_{i}}}\right) \right) \right) \right. \\
& \left. + I(y_{i,t}=2) \times \log \left(\frac{1}{L} \sum_{l=1}^L \sum_{k=1}^{K_i^{(l)}} w_{ik}^{(l)} \left( \Phi \left(\frac{2 - (\beta_{0i}^{(l,k)} + \bfX_{i,t} \bfbeta_{i}^{(l)} + \bfH_{i,t} \bfgamma_{i}^{(l)})}{1/\sqrt{\tau^{2^{(l,k)}}_{i}}}\right) - \Phi \left(\frac{1 - (\beta_{0i}^{(l,k)} + \bfX_{i,t} \bfbeta_{i}^{(l)} + \bfH_{i,t} \bfgamma_{i}^{(l)})}{1/\sqrt{\tau^{2^{(l,k)}}_{i}}}\right) \right) \right) \right. \\[3mm]
& \hspace{7cm} \fivevdots \\[3mm]
& \left. + I(y_{i,t}=J) \times \log \left(\frac{1}{L} \sum_{l=1}^L \sum_{k=1}^{K_i^{(l)}} w_{ik}^{(l)} \left( 1 - \Phi \left(\frac{J-1 - (\beta_{0i}^{(l,k)} + \bfX_{i,t} \bfbeta_{i}^{(l)} + \bfH_{i,t} \bfgamma_{i}^{(l)})}{1/\sqrt{\tau^{2^{(l,k)}}_{i}}}\right) \right) \right) \right),
\end{aligned}
\end{equation}
}
where the term $w_{ik}^{(l)}$ denotes the posterior weight of the $k^{\text{th}}$ mixture component in the $l^{\text{th}}$ iteration, which arises from the stick-breaking representation of the Dirichlet process.

%% file: 4-results.tex
\subsection{BNP and BP Models without a Temporal Random Effect}
Here, we fit both the  BP and BNP models without a temporal random effect to the drought data, as specified in (\ref{eq:bp}) and (\ref{eq:bnp}). The MCMC sampling algorithms for each location were implemented in parallel using an array job configuration on the DEAC high-performance computing cluster at Wake Forest University (\cite{WakeHPC}). Each MCMC chain was run for $100{,}000$ iterations, discarding the first $50{,}000$ iterations as burn-in and applying a thinning interval of $10$. All MCMC algorithms were implemented using NIMBLE, an R-based system that compiles hierarchical models and MCMC algorithms into C++ \cite[see][]{nimble-article:2017,nimble-software:2024,nimble-manual:2024}. The computational time for a single grid cell was approximately $10$ minutes for the BNP model without basis functions and $7$ minutes for the BP model without basis functions when running on one core on a single node. This duration includes the time required to load the drought data, build and compile the model in NIMBLE, run the MCMC for $100{,}000$ iterations, and compute the log score.

NIMBLE supports the specification of BNP models, including hierarchical formulations involving Dirichlet process mixture models (DPMM). These models can be expressed in different, yet equivalent, ways, each leading NIMBLE to apply different sampling algorithms for the model. In particular, NIMBLE provides built-in functionality for fitting DPMMs using either the Chinese Restaurant Process (CRP) representation or the stick-breaking representation, along with the necessary structural definitions and MCMC algorithms required for their implementation \cite[see,][]{nimble-manual:2024}. In this study, we adopt the CRP representation in NIMBLE to fit our BNP models. When the model is specified using the CRP representation in NIMBLE, a collapsed Gibbs sampler (\cite{neal2000markov}) is automatically employed. 

\subsection{BNP and BP Models with a Temporal Random Effect}
In this section, we fit the BP and BNP models with a temporal random effect, as specified in (\ref{eq:bp-2q}) and (\ref{eq:bnp-2q}), to account for temporal autocorrelation and to capture periodic or seasonal patterns present in the drought data.  

We evaluate the performance of five model configurations. First, we consider a BP model with a temporal random effect represented using $Q=10$ pairs of Fourier basis functions, hereafter referred to as \texttt{BP20}. Subsequently, we fit four BNP models with a temporal random effect represented using $Q = 7, 8, 9,$ and $10$ pairs of Fourier basis functions, corresponding to \texttt{BNP14}, \texttt{BNP16}, \texttt{BNP18}, and \texttt{BNP20}, respectively. The MCMC algorithms were again executed in parallel using an array job configuration on the DEAC high-performance computing cluster (\cite{WakeHPC}). Each MCMC chain was run in NIMBLE for a total of $100{,}000$ iterations, with the first $50{,}000$ iterations discarded as burn-in and a thinning interval of $10$ applied to reduce autocorrelation. The run time for a single grid cell was approximately one hour for the BNP models with a temporal random effect and about ten minutes for the BP model with a temporal random effect, utilizing one core on a single node. This time includes loading the drought data, constructing and compiling the model in NIMBLE, running the MCMC for $100{,}000$ iterations, and computing the log score.

\subsection{Model Comparison}
In this section, we present a comparison of the BNP and BP models, utilizing the log score as defined in equations (\ref{eq:log-calc-bnp}) and (\ref{eq:log-calc-bp}). Table \ref{tab:logscore-sum} presents the cumulative sum of log scores across all spatial locations for the BNP and BP models without a temporal random effect. These results indicate that the BNP model, with a higher total log score of $-3236.973$, demonstrates superior overall predictive performance compared to the BP model, which achieved a total log score of $-3317.264$. 

\begin{table}[h]
\centering
\setlength{\tabcolsep}{18pt} 
\begin{tabular}{l c}
\toprule
\textbf{Model} & \textbf{Sum of Log Scores} \\
\midrule
\texttt{BNP} & $-3236.973$ \\
\texttt{BP}  & $-3317.264$ \\
\bottomrule
\end{tabular}
\caption{Sum of log scores across spatial locations for the \texttt{BNP} and \texttt{BP} models without a temporal random effect.}
\label{tab:logscore-sum}
\end{table}

\begin{figure}[h]
    \centering
    \includegraphics[width=0.99\textwidth]{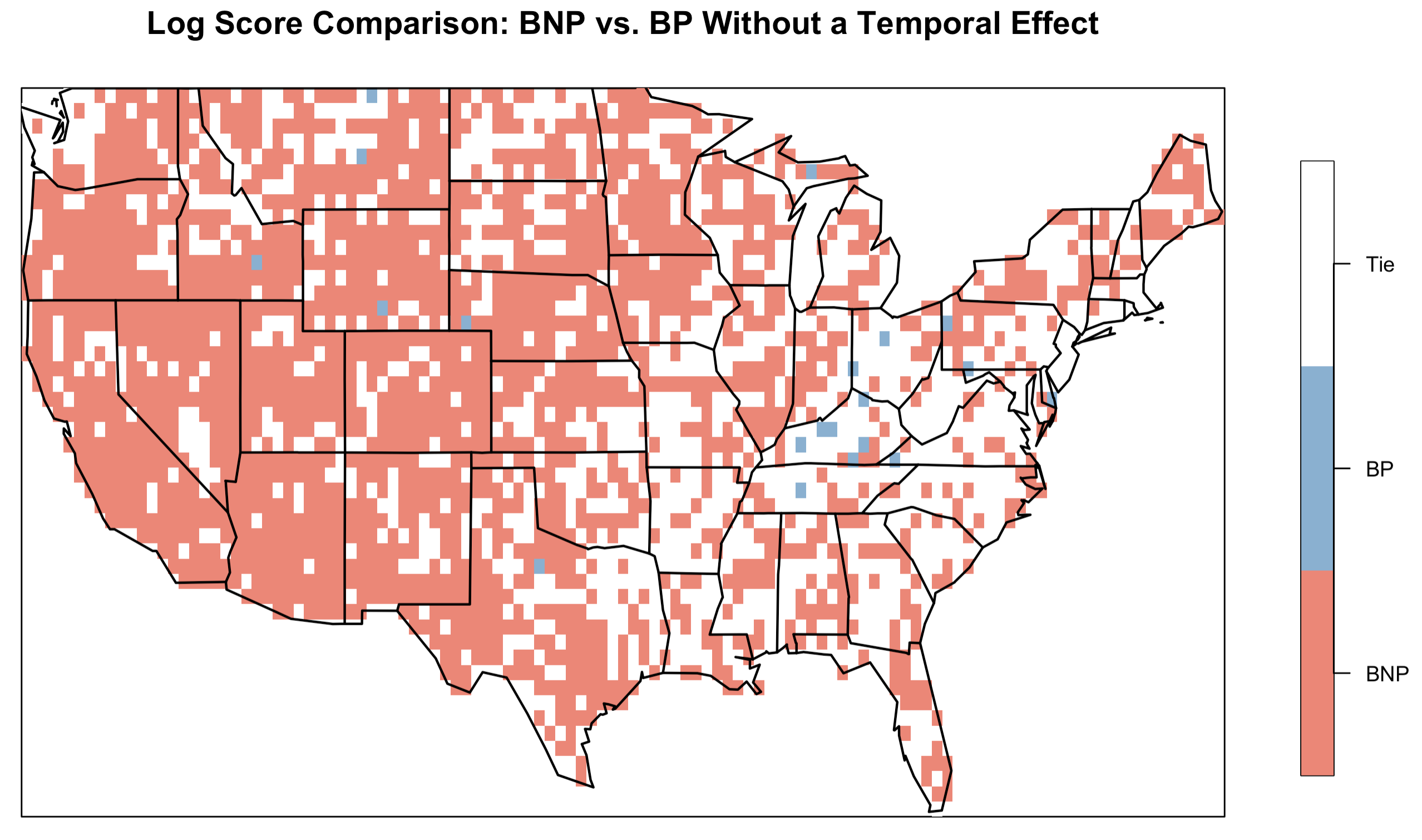}
    \caption{Spatial comparison of log scores for the \texttt{BNP} and \texttt{BP} models without a temporal random effect, highlighting the better-performing model at each location. Locations with the absolute log score difference below $0.001$ are labeled as \texttt{Tie}.}
    \label{fig:bnp-bp-logscores}
\end{figure}

To further illustrate these findings, a comparison map, shown in Figure \ref{fig:bnp-bp-logscores}, is constructed to visualize the log score results for the BNP and BP models without a temporal random effect across locations, highlighting the model that exhibits superior predictive performance at each location. At locations where the absolute value of the difference in log scores between the two models is less than $0.001$, the predictive performance is considered comparable, and such locations are represented with white shading to denote a "\texttt{Tie}". 

Additionally, Table \ref{tab:bnp-bp-perc} summarizes the percentage of spatial locations where the BNP, BP, or neither model (\texttt{Tie}) achieved the highest log score, based on the comparison shown in Figure \ref{fig:bnp-bp-logscores}. The results clearly indicate that the BNP model exhibits superior predictive performance across the majority of locations, attaining the highest log score in $61.5\%$ of the spatial domain. In $37.9\%$ of locations, the log score differences between the two models are negligible, and these locations are designated as \texttt{Tie}. Overall, these findings suggest the enhanced predictive capabilities of the BNP model for modeling  drought data compared to the BP model.

\begin{table}[h]
\centering
\setlength{\tabcolsep}{20pt} 
\begin{tabular}{l c}
\toprule
\textbf{Model} & \textbf{Percentage} \\
\midrule
\texttt{BNP}   & 61.5\% \\
\texttt{BP}    & 0.6\%  \\
\texttt{Tie}   & 37.9\% \\
\bottomrule
\end{tabular}
\caption{Percentage of spatial locations where the \texttt{BNP}, \texttt{BP}, or neither model (\texttt{Tie}) achieved the highest log score in the comparison shown in Figure~\ref{fig:bnp-bp-logscores}.}
\label{tab:bnp-bp-perc}
\end{table}

A separate comparison is presented in Table \ref{tab:logscore-bnp20-bp20}, which reports the sum of log scores across spatial locations for the \texttt{BNP14}, \texttt{BNP16}, \texttt{BNP18}, \texttt{BNP20}, and \texttt{BP20} models with a temporal random effect. Among these models, \texttt{BNP20} achieves the highest cumulative log score of $-1390.375$, indicating superior overall predictive performance. Notably, both \texttt{BNP20} and \texttt{BNP18} outperform the \texttt{BP20} model, which yields a total log score of $-1531.512$. In addition, a comparison of Tables~\ref{tab:logscore-bnp20-bp20} and~\ref{tab:logscore-sum} shows that the inclusion of a temporal random effect substantially improves predictive performance, as all temporal models achieve much higher cumulative log scores than the non-temporal models. These results highlight the effectiveness of Bayesian nonparametric modeling and the use of a temporal random effect, implemented through Fourier basis functions, in capturing temporal variation in drought data across spatial locations.

\begin{table}[h]
\centering
\setlength{\tabcolsep}{20pt} 
\begin{tabular}{l c}
\toprule
\textbf{Model} & \textbf{Sum of Log Scores} \\
\midrule
\texttt{BNP14} & $-1685.598$ \\
\texttt{BNP16} & $-1590.742$ \\
\texttt{BNP18} & $-1481.109$ \\
\texttt{BNP20} & $-1390.375$ \\
\texttt{BP20}  & $-1531.512$ \\
\bottomrule
\end{tabular}
\caption{Sum of log scores across spatial locations for \texttt{BNP14}, \texttt{BNP16}, \texttt{BNP18}, \texttt{BNP20}, and \texttt{BP20} models with a temporal random effect.}
\label{tab:logscore-bnp20-bp20}
\end{table}

\begin{figure}[h]
    \centering
    \includegraphics[width=0.99\textwidth]{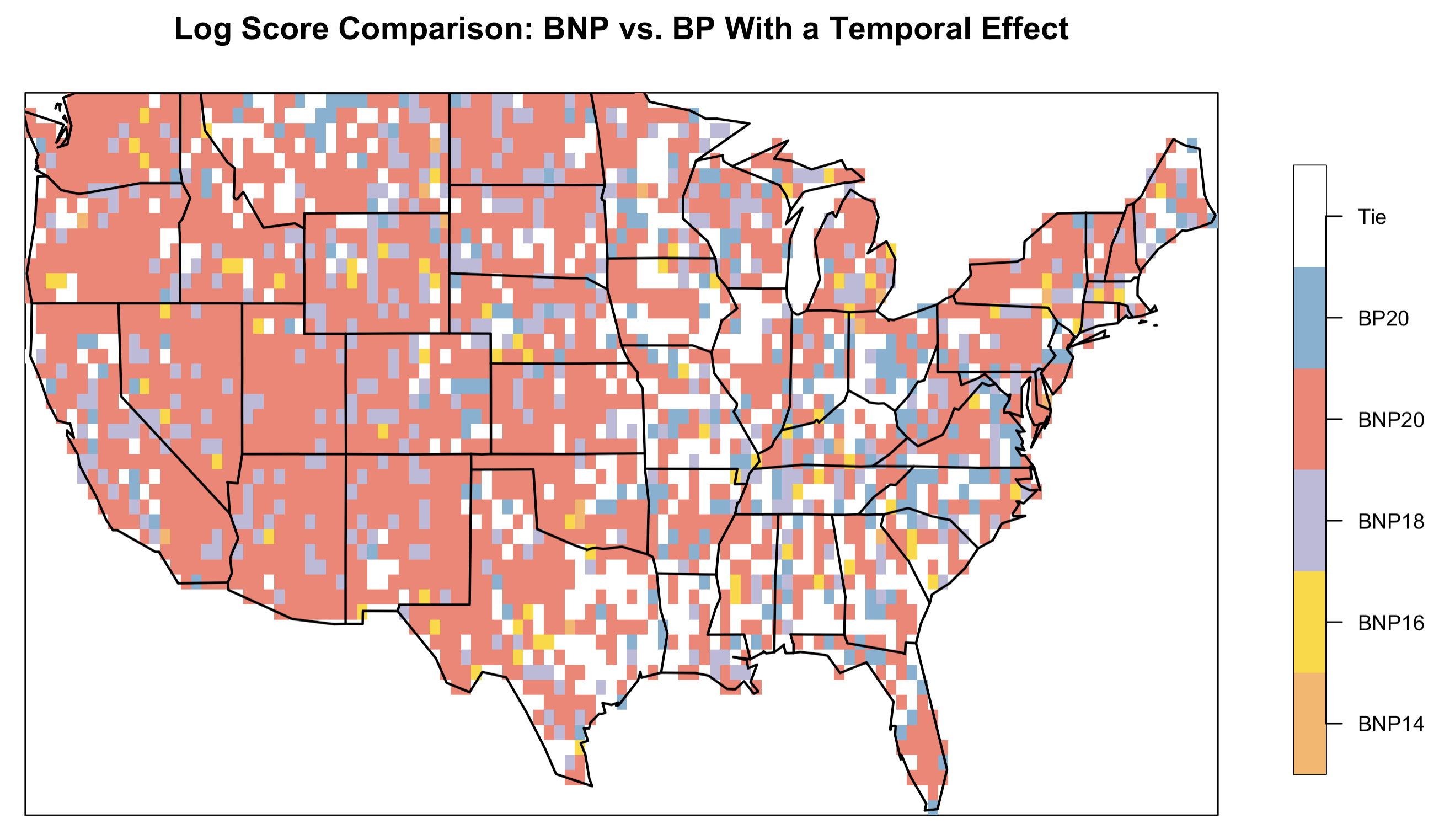}
    \caption{Map displaying location-wise log score comparisons among the \texttt{BNP14}, \texttt{BNP16}, \texttt{BNP18}, \texttt{BNP20}, and \texttt{BP20} models fitted with a temporal random effect. A \texttt{Tie} is assigned where the absolute difference between the highest log score among the BNP models and that of the \texttt{BP20} model is less than $0.001$.}
    \label{fig:bnp-f-logscores}
\end{figure}

This comparison is further visualized in Figure \ref{fig:bnp-f-logscores}, which displays the log score comparison map for \texttt{BNP14}, \texttt{BNP16}, \texttt{BNP18}, \texttt{BNP20}, and \texttt{BP20} models. The figure shows the model with the highest log score at each location. Locations where the absolute difference between the maximum log score among the BNP models (\texttt{BNP20}, \texttt{BNP18}, \texttt{BNP16}, and \texttt{BNP14}) and the log score of the \texttt{BP20} model is less than $0.001$ are labeled as "\texttt{Tie}", indicating comparable predictive performance.    

Finally, Table \ref{tab:bnp20-bp20-perc} summarizes the percentage of spatial locations at which each model achieved the highest log score in the spatial comparison presented in Figure~\ref{fig:bnp-f-logscores}. The results indicate that the \texttt{BNP20} model performs best across the majority of locations, attaining the highest log score in $59.4\%$ of the spatial domain. Moreover, the combined dominance of the BNP models (\texttt{BNP20}, \texttt{BNP18}, \texttt{BNP16}, and \texttt{BNP14}) collectively exceeds the \texttt{BP20} model. In $18.6\%$ of locations, the log score differences between models are negligible, resulting in a \texttt{Tie}.

\begin{table}[ht!]
\centering
\setlength{\tabcolsep}{20pt} 
\begin{tabular}{l c}
\toprule
\textbf{Model} & \textbf{Percentage} \\
\midrule
\texttt{BNP20} & 59.4\% \\
\texttt{BNP18} & 12.0\%  \\
\texttt{BNP16} & 2.3\%  \\
\texttt{BNP14} & 0.5\%  \\
\texttt{BP20}  & 7.2\%  \\
\texttt{Tie}   & 18.6\% \\
\bottomrule
\end{tabular}
\caption{Percentage of spatial locations where each model achieved the highest log score in the comparison shown in Figure~\ref{fig:bnp-f-logscores}.}
\label{tab:bnp20-bp20-perc}
\end{table}

\subsection{Exploring Performance Differences Between \texttt{BNP20} and \texttt{BP20}}
The \texttt{BNP20} model offers greater flexibility than the \texttt{BP20} model by clustering time periods within each spatial location and allowing the time-specific intercept and precision parameters $(\beta_{0,it},\tau^2_{it})$ to vary across these clusters, as induced by the Dirichlet process prior. However, this flexibility does not guarantee that \texttt{BNP20} will consistently outperform \texttt{BP20}. The relative performance of the two models depends on the underlying complexity and structure of the drought data at each spatial location. To investigate this further, we examine three representative grid locations labeled as: EE12 (longitude: $-119.25^\circ$, latitude: $34.75^\circ$), where \texttt{BNP20} performs better; U95 (longitude: $-77.75^\circ$, latitude: $39.75^\circ$), where \texttt{BP20} performs slightly better; and TT53 (longitude: $-98.75^\circ$, latitude: $27.25^\circ$), where both models tie.

\begin{figure}[ht]
    \centering
    \includegraphics[width=0.99\textwidth]{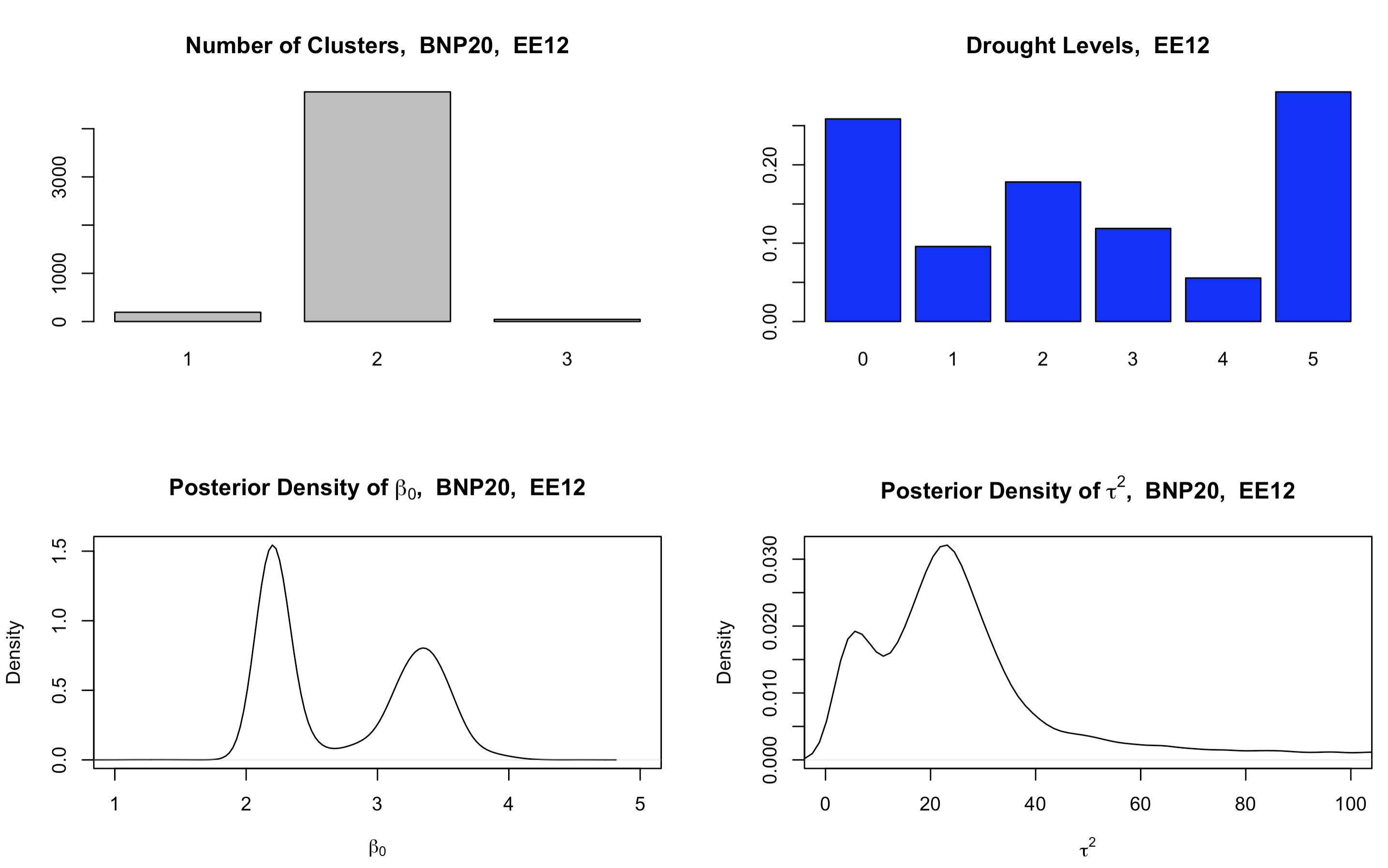}
    \includegraphics[width=0.99\textwidth]{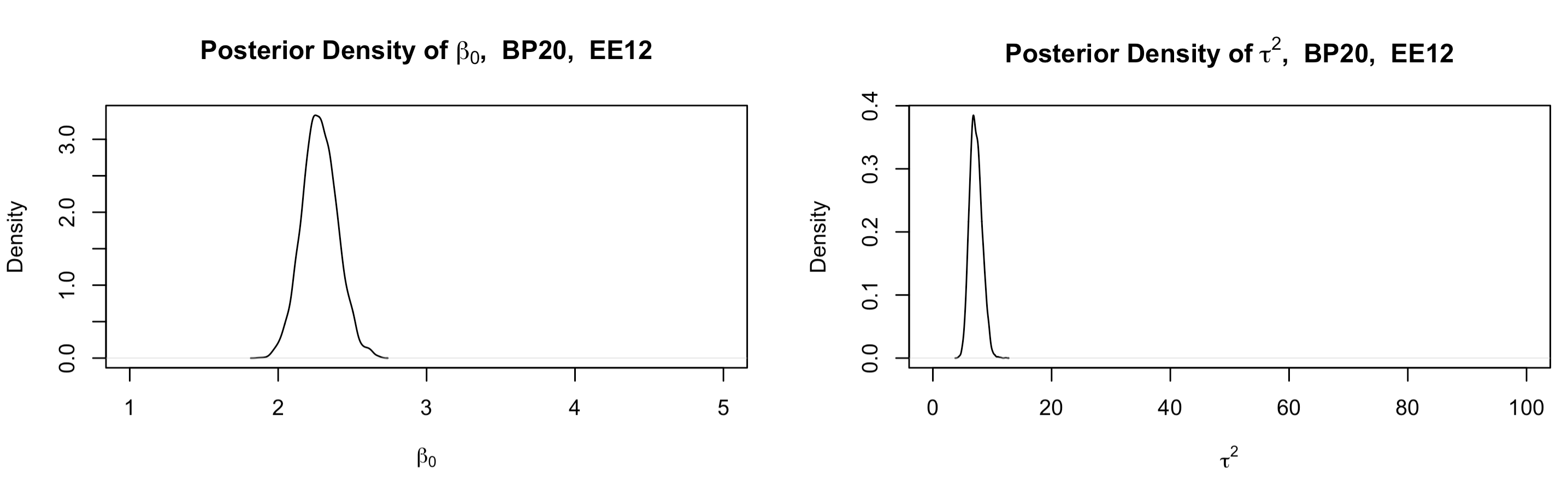}
    \caption{Posterior plots and drought level distribution for location EE12. Top-left: \texttt{BNP20} posterior distribution of the number of clusters; Top-right: distribution of observed drought levels; Middle: \texttt{BNP20} posterior densities of $\beta_0$ and $\tau^2$; Bottom: \texttt{BP20} posterior densities of $\beta_0$ and $\tau^2$.}
    \label{fig:EE12}
\end{figure}

For locations with a multimodal and complex data structure, the \texttt{BNP20} model's flexibility provides a significant advantage. For example, Figure \ref{fig:EE12} presents the posterior distribution of the number of clusters under \texttt{BNP20}, the bar plot of observed drought levels, the posterior densities of the intercept ($\beta_0$) and precision ($\tau^2$) under \texttt{BNP20}, and the corresponding posterior densities under \texttt{BP20}, all for the EE12 location. The drought level distribution at EE12 is bimodal. The posterior analysis for \texttt{BNP20} at this location demonstrates a clear preference for two clusters and bimodal posterior densities for both $\beta_0$ and $\tau^2$. In contrast, the \texttt{BP20} model attempts to fit this complex distribution with a single component, resulting in a poorer fit and a lower log score. The posterior means and standard deviations of the parameters $\beta_0$, $\tau^2$, $\beta_{\texttt{evp}}$, and $\beta_{\texttt{stream28}}$ for location EE12 are presented in Table \ref{tab:EE12_post} in the Appendix. 
 
\begin{figure}[ht]
    \centering
    \includegraphics[width=0.99\textwidth]{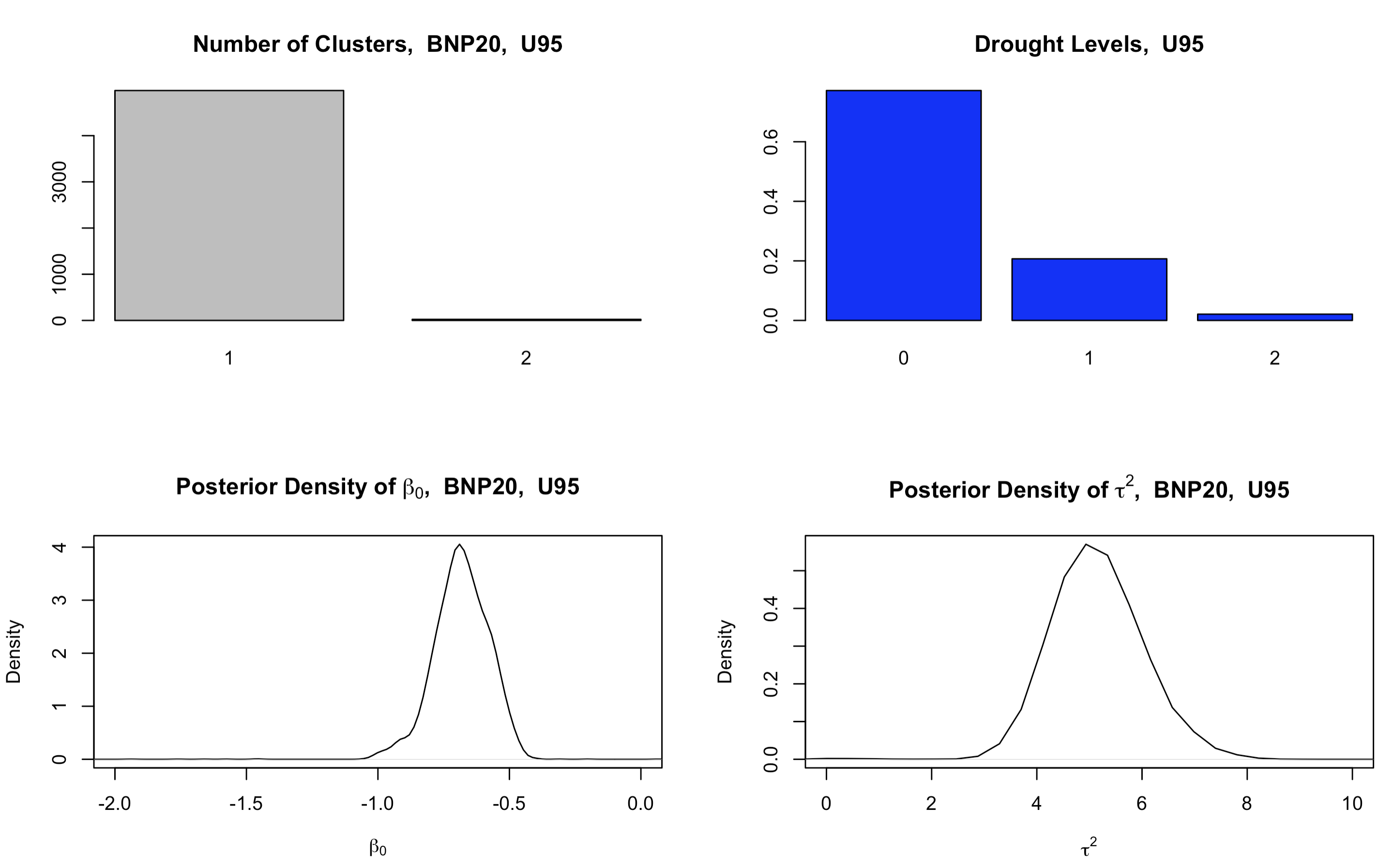}
    \includegraphics[width=0.99\textwidth]{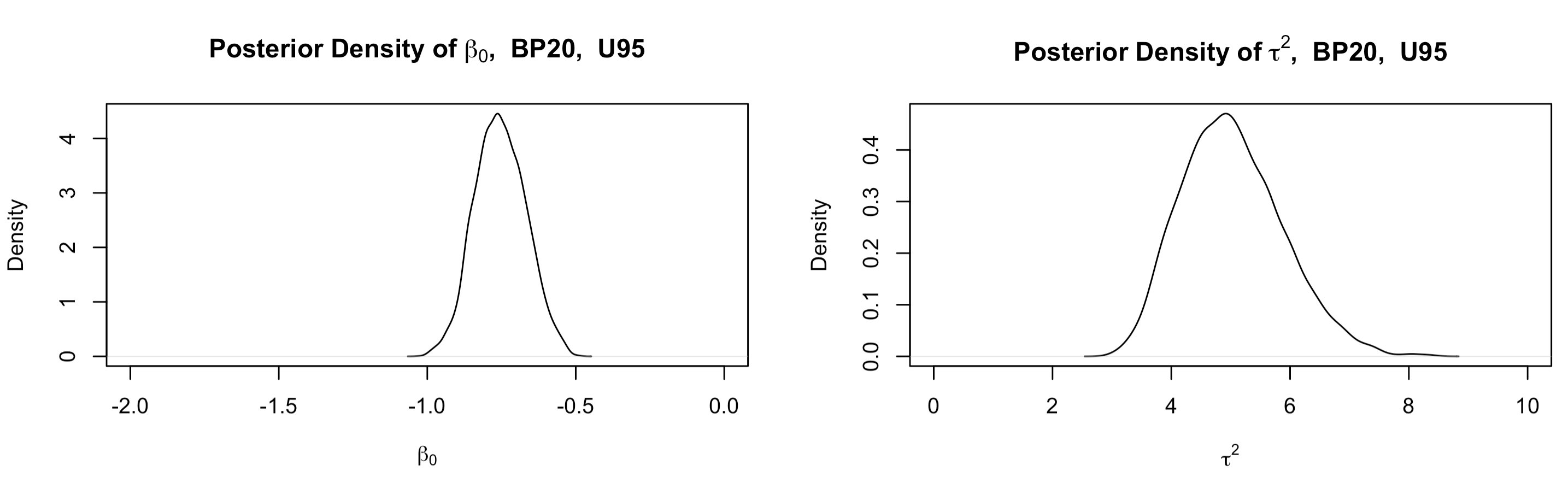}
    \caption{Posterior plots and drought level distribution for location U95. Top-left: \texttt{BNP20} posterior distribution of the number of clusters; Top-right: distribution of observed drought levels; Middle: \texttt{BNP20} posterior densities of $\beta_0$ and $\tau^2$; Bottom: \texttt{BP20} posterior densities of $\beta_0$ and $\tau^2$.}
    \label{fig:U95}
\end{figure}

When the drought data distribution is simple and unimodal, the flexibility of the \texttt{BNP20} model offers no significant advantage, leading to a performance that is either a \texttt{Tie} or slightly inferior to the \texttt{BP20} model. For instance, at location U95 (Figure \ref{fig:U95}), the drought levels are concentrated in the lower categories, and the posterior clustering under \texttt{BNP20} collapses to a single cluster. As a result, \texttt{BNP20} behaves similarly to a parametric model, but \texttt{BP20}'s simpler structure allows for more stable posterior estimates and a slightly better log score. 

\begin{figure}[ht]
    \centering
    \includegraphics[width=0.99\textwidth]{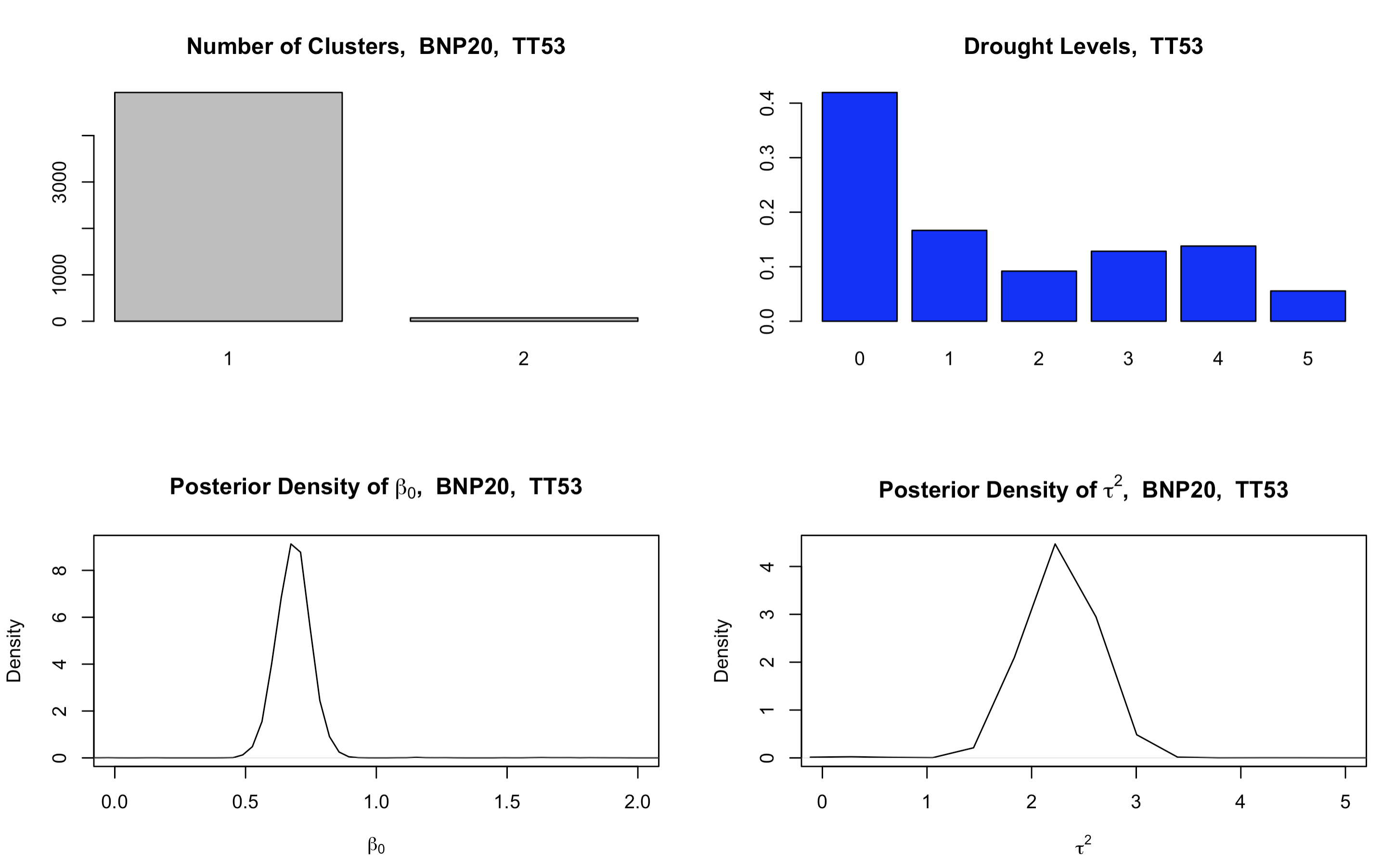}
    \includegraphics[width=0.99\textwidth]{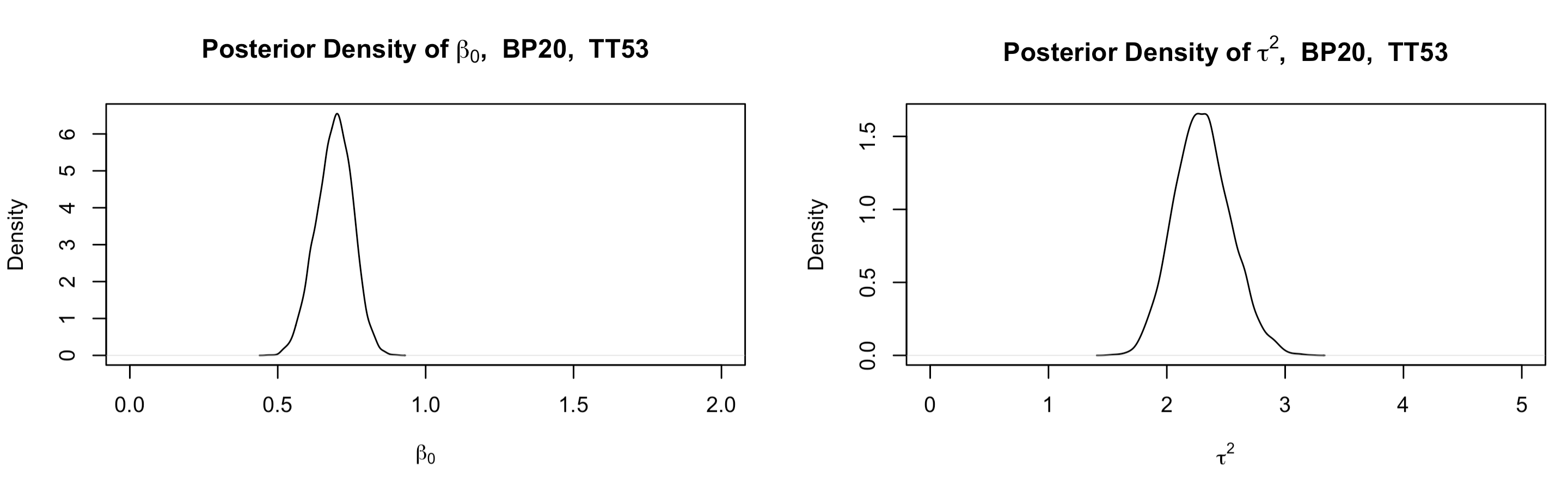}
    \caption{Posterior plots and drought level distribution for location TT53. Top-left: \texttt{BNP20} posterior distribution of the number of clusters; Top-right: distribution of observed drought levels; Middle: \texttt{BNP20} posterior densities of $\beta_0$ and $\tau^2$; Bottom: \texttt{BP20} posterior densities of $\beta_0$ and $\tau^2$.}
    \label{fig:TT53}
\end{figure}

A similar situation is observed at location TT53 (Figure \ref{fig:TT53}), where both models perform equally well in terms of log score. Like U95, \texttt{BNP20} identifies only a single cluster of time periods, resulting in near-identical behavior to \texttt{BP20}. Tables \ref{tab:U95_post} and \ref{tab:TT53_post} in the Appendix summarize the posterior means and standard deviations of the parameters $\beta_0$, $\tau^2$, $\beta_{\texttt{evp}}$, and $\beta_{\texttt{stream28}}$ for locations U95 and TT53. It is important to note that the \texttt{BNP20} model yields higher posterior standard deviations for the parameters $\beta_0$ and $\tau^2$, which are assigned DP priors, compared to the \texttt{BP20} model.

A summary of these results is provided in Table \ref{tab:logscore-comparison}. These examples suggest that \texttt{BNP20} only outperforms \texttt{BP20} when the \texttt{BNP20} model finds meaningful heterogeneity in the data, by forming multiple mixture components. When only a single cluster is favored, the \texttt{BNP20} model reduces to a parametric form, offering little benefit over \texttt{BP20} and sometimes performing slightly worse due to increased posterior uncertainty for the parameters with DP priors ($\beta_0$ and $\tau^2$).

\begin{table}[H]
\centering
\begin{tabular}{lccc}
\toprule
Grid & Best Model & Log Score Difference (\texttt{BNP20} - \texttt{BP20}) & \# of Clusters in \texttt{BNP20} \\
\midrule
EE12 & \texttt{BNP20} & $+0.034$ & $ > 1$ \\
U95  & \texttt{BP20}  & $-0.002$ & $1$ \\
TT53 & \texttt{Tie}            & $\approx 0$ & $1$ \\
\bottomrule
\end{tabular}
\caption{Log score comparison across models for selected locations}
\label{tab:logscore-comparison}
\end{table}


\clearpage
\subsection{Posterior Uncertainty in Covariate Effects Across Models}

Table \ref{tab:mean_post_sd} presents the mean of posterior standard deviations of the regression coefficients $\beta_{\texttt{evp}}$ and $\beta_{\texttt{stream28}}$ across all spatial locations for the \texttt{BP20} and \texttt{BNP} models with varying numbers of basis functions. These mean posterior standard deviations provide a measure of uncertainty in the estimated covariate effects, $\beta_{\texttt{evp}}$ and $\beta_{\texttt{stream28}}$, which are modeled with parametric priors rather than DP priors.

As shown in Table \ref{tab:mean_post_sd}, the \texttt{BNP} models consistently produce smaller mean posterior standard deviations compared to the parametric \texttt{BP20} model. The \texttt{BP20} model exhibits the highest level of posterior uncertainty, with mean posterior standard deviations of $0.13$ for $\beta_{\texttt{evp}}$ and $0.08$ for $\beta_{\texttt{stream28}}$. In comparison, the \texttt{BNP20} model reduces the uncertainty for both covariates. A further reduction is observed as the number of basis functions is decreased in the \texttt{BNP} models. Specifically, the \texttt{BNP18}, \texttt{BNP16}, and \texttt{BNP14} models yield nearly identical mean posterior standard deviations for both coefficients, around $0.07$ for $\beta_{\texttt{evp}}$ and $0.06$ for $\beta_{\texttt{stream28}}$.

These results indicate that the \texttt{BNP} models generally have greater confidence in the estimates of the covariate effects, as evidenced by their lower posterior uncertainty. This suggests a better fit to the data and more stable parameter estimates when using the more flexible BNP approach.

\begin{table}[h!]
\centering
\vspace{0.3cm}
\begin{tabular}{lcc}
\toprule
Model & Mean posterior SD of $\beta_{\texttt{evp}}$ & Mean posterior SD of $\beta_{\texttt{stream28}}$ \\
\midrule
\texttt{BP20} & 0.13 & 0.08 \\
\texttt{BNP20} & 0.11 & 0.07 \\
\texttt{BNP18} & 0.07 & 0.06 \\
\texttt{BNP16} & 0.07 & 0.06 \\
\texttt{BNP14} & 0.07 & 0.06 \\
\bottomrule
\end{tabular}
\caption{Mean of posterior standard deviations for parameters $\beta_{\texttt{evp}}$ and $\beta_{\texttt{stream28}}$ across all locations for different models.}
\label{tab:mean_post_sd}
\end{table}

%% file: 5-discussion.tex
In this paper, we showed that a Bayesian nonparametric model for ordinal drought data achieves enhanced model flexibility in assigning probabilities to the levels of the response variable.  This extends model flexibility for ordinal models that rely on a latent continuous variable, even when the cutoff parameter $\bfalpha$ is fixed for computational reasons.  In our application, we achieved enhanced model flexibility by assigning a Dirichlet process prior to the joint mixing distribution of the time-specific intercept and precision parameters $(\beta_{0,it},\tau^2_{it})$, while assigning all other parameters parametric priors. The Dirichlet process induces clustering among time periods within each spatial location, with time periods assigned to the same cluster sharing the same intercept and precision parameters. Our BNP model permits a mixture of Gaussian distributions for the latent variable, each component of which has a different intercept and/or precision.  Results showed that even with only two Gaussian distributions constituting the mixture, the model can flexibly assign probability across the $J+1=6$ ordinal levels in this example and provide a superior model fit as measured by the log-scoring metric. In addition, we demonstrated that including a temporal random effect, represented using Fourier basis functions, to capture temporal dependence, further improves predictive performance.

Our BNP model outperformed comparable Bayesian parametric models when evaluated by the log score. In many cases, BNP showed a notably superior fit; in other cases, the result was essentially a tie between BNP and BP, often arising from the BNP model favoring a single cluster of time periods and therefore fitting a Gaussian mixture with only one component for the latent variable. This overall pattern of results held after we included a temporal random effect via Fourier basis functions in both models. When comparing the BNP and BP models with a temporal random effect, BNP frequently achieved better performance, sometimes with fewer basis functions, and often tied in performance for other cases.  

The primary motivation for this paper was to achieve higher model flexibility in assigning probabilities to the $J+1$ ordinal categories through a latent Gaussian mixture while retaining a fixed $\bfalpha$.  We showed this with only two parameters getting BNP priors.  A consequence is that the posterior distribution of other parameters for fixed effects changes in response to the BNP prior, and so inference on other fixed effects changes in consequence.  Enhanced model flexibility through BNP priors on some parameters can result in more precise credible intervals for other parameters, as shown in our application.

One open question is whether a BNP prior for $\bfalpha$ is feasible, and if this might assist with model flexibility and/or known challenges with computational cost and mixing.  The ordering that $\alpha_0 < \alpha_1 < \cdots < \alpha_{J} < \alpha_{J+1}$ would pose an additional challenge, but a transformation used in \cite{erhardt2024spatio} could ensure that ordering is preserved.  Another open question is how spatial dependence could be captured alongside the use of BNP priors, and how the resulting model complexity would lead to a growth in computational complexity. In this work, we captured temporal dependence using Fourier basis functions; however, extending this framework to simultaneously model spatial dependence remains a challenging direction for future research. The BNP priors unquestionably increase the computational cost and model complexity, as spatial dependence would as well. How those costs scale with data is an open question.

%% file: 6-appendix.tex
\begin{table}[H] 
\centering
\caption{Posterior mean and standard deviation for the main parameters in \texttt{BNP20} and \texttt{BP20} for location \textbf{EE12}.}
\label{tab:EE12_post}
\vspace{0.3cm}
\begin{tabular}{lcccc}
\hline
 & \multicolumn{2}{c}{\texttt{BNP20}} & \multicolumn{2}{c}{\texttt{BP20}} \\
\cline{2-3} \cline{4-5}
Parameter & Mean & SD & Mean & SD \\
\hline
$\beta_0$ & 2.73 & 0.60 & 2.28 & 0.12 \\
$\tau^2$  & 39.50 & 54.26 & 7.25 & 1.02 \\
$\beta_{\texttt{evp}}$ & 0.47 & 0.18 & 0.55 & 0.21 \\
$\beta_{\texttt{stream28}}$ & -0.39 & 0.07 & -0.43 & 0.07 \\
\hline
\end{tabular}
\end{table}

\begin{table}[H] 
\centering
\caption{Posterior mean and standard deviation for the main parameters in \texttt{BNP20} and \texttt{BP20} for location \textbf{U95}.}
\label{tab:U95_post}
\vspace{0.3cm}
\begin{tabular}{lcccc}
\hline
 & \multicolumn{2}{c}{\texttt{BNP20}} & \multicolumn{2}{c}{\texttt{BP20}} \\
\cline{2-3} \cline{4-5}
Parameter & Mean & SD & Mean & SD \\
\hline
$\beta_0$ & -0.68 & 0.18 & -0.75 & 0.09 \\
$\tau^2$  & 5.22 & 3.16 & 5.03 & 0.85 \\
$\beta_{\texttt{evp}}$ & -0.18 & 0.09 & -0.21 & 0.10 \\
$\beta_{\texttt{stream28}}$ & -0.82 & 0.09 & -0.85 & 0.09 \\
\hline
\end{tabular}
\end{table}

\begin{table}[H] 
\centering
\caption{Posterior mean and standard deviation for the main parameters in \texttt{BNP20} and \texttt{BP20} for location \textbf{TT53}.}
\label{tab:TT53_post}
\vspace{0.3cm}
\begin{tabular}{lcccc}
\hline
 & \multicolumn{2}{c}{\texttt{BNP20}} & \multicolumn{2}{c}{\texttt{BP20}} \\
\cline{2-3} \cline{4-5}
Parameter & Mean & SD & Mean & SD \\
\hline
$\beta_0$ & 0.67 & 0.30 & 0.69 & 0.06 \\
$\tau^2$  & 2.50 & 4.30 & 2.30 & 0.24 \\
$\beta_{\texttt{evp}}$ & -0.47 & 0.12 & -0.46 & 0.12 \\
$\beta_{\texttt{stream28}}$ & -0.63 & 0.13 & -0.62 & 0.13 \\
\hline
\end{tabular}
\end{table}

%% file: ref.bib
@article{erhardt2024homogenized,
  title={Homogenized gridded dataset for drought and hydrometeorological modeling for the continental {U}nited {S}tates},
  author={Erhardt, Robert and Di Vittorio, Courtney A and Hepler, Staci A and Lowman, Lauren EL and Wei, Wendy},
  journal={Scientific Data},
  volume={11},
  number={1},
  pages={375},
  year={2024},
  publisher={Nature Publishing Group UK London}
}

@book{banerjee2014hierarchical,
  title={Hierarchical Modeling and Analysis for Spatial Data},
  author={Banerjee, Sudipto and Carlin, Bradley P and Gelfand, Alan E},
  year={2014},
  publisher={CRC Press}
}

@article{agresti2014some,
  title={Some remarks on latent variable models in categorical data analysis},
  author={Agresti, Alan and Kateri, Maria},
  journal={Communications in Statistics-Theory and Methods},
  volume={43},
  number={4},
  pages={801--814},
  year={2014},
  publisher={Taylor \& Francis}
}

@book{agresti2010analysis,
  title={Analysis of ordinal categorical data},
  author={Agresti, Alan},
  year={2010},
  publisher={John Wiley \& Sons}
}

@article{erhardt2024spatio,
  title={Spatio-temporal forecasting for the US Drought Monitor},
  author={Erhardt, Robert and Hepler, Staci and Wolodkin, Daniel and Greene, Andy},
  journal={Journal of the Royal Statistical Society Series C: Applied Statistics},
  volume={73},
  number={5},
  pages={1203--1220},
  year={2024},
  publisher={Oxford University Press UK}
}

@article{hepler-erhardt2025arxiv,
      title={{Two-stage MCMC for Fast Bayesian Inference of Large Spatio-temporal Ordinal Data, with Application to US Drought}}, 
      author = {Hepler, Staci and Erhardt, Rob},
      year={2025},
      eprint={2505.24594},
      journal = {arXiv preprint arXiv:2505.24594},
      archivePrefix={arXiv},
      primaryClass={stat.ME},
      url={https://arxiv.org/abs/2505.24594}, 
}

@misc{WakeHPC,
  doi       = {10.57682/G13Z-2362},
  url       = {https://hpc.wfu.edu},
  author    = {{Information Systems and Wake Forest University}},
  title     = {{WFU High Performance Computing Facility}},
  publisher = {Wake Forest University},
  year      = {2021}
}

@article{ferguson1973,
author = "Ferguson, Thomas S.",
doi = "10.1214/aos/1176342360",
fjournal = "Annals of Statistics",
journal = {The Annals of Statistics},
month = "03",
number = "2",
pages = "209--230",
publisher = "The Institute of Mathematical Statistics",
title = "A {B}ayesian Analysis of Some Nonparametric Problems",
volume = "1",
year = "1973"
}

@article{sethuraman1994constructive,
  title={A constructive definition of {D}irichlet priors},
  author={Sethuraman, Jayaram},
  journal={Statistica Sinica},
  pages={639--650},
  year={1994},
  publisher={JSTOR}
}

@article{schliep2015data,
  title={Data augmentation and parameter expansion for independent or spatially correlated ordinal data},
  author={Schliep, Erin M and Hoeting, Jennifer A},
  journal={Computational Statistics \& Data Analysis},
  volume={90},
  pages={1--14},
  year={2015},
  publisher={Elsevier}
}

@article{nimble-article:2017, 
  author = {{de Valpine}, P. and Turek, D. and Paciorek, C.J. and Anderson-Bergman, C. and {Temple Lang}, D. and Bodik, R.}, 
  title = {Programming with models: writing statistical algorithms for general model structures with {NIMBLE}},
  year = {2017}, 
  journal = {Journal of Computational and Graphical Statistics},
  volume = 26,
  pages = {403-413},
  doi = {10.1080/10618600.2016.1172487}
}

@misc{nimble-software:2024,
  author = {{de Valpine}, P. and Paciorek, C. and Turek, D. and Michaud, N. and Anderson-Bergman, C. and Obermeyer, F. and Wehrhahn Cortes, C. and Rodr{\'i}guez, A. and {Temple Lang}, D. and Zhang, W. and Paganin, S. and Hug, J. and van Dam-Bates, P.}, 
  year = {2024},
  title = {NIMBLE: MCMC, Particle Filtering, and Programmable Hierarchical Modeling},
  version = {1.2.1},
  note = {{R} package version 1.2.1},
  doi = {10.5281/zenodo.1211190},
  url = {https://cran.r-project.org/package=nimble}
}

@manual{nimble-manual:2024,
  author = {{de Valpine}, P. and Paciorek, C. and Turek, D. and Michaud, N. and Anderson-Bergman, C. and Obermeyer, F. and Wehrhahn Cortes, C. and Rodr{\'i}guez, A. and {Temple Lang}, D. and Zhang, W. and Paganin, S. and Hug, J. and van Dam-Bates, P.}, 
  year = {2024}, 
  title = {NIMBLE User Manual}, 
  version = {1.2.1}, 
  note = {{R} package manual version 1.2.1},
  doi = {10.5281/zenodo.1211190}, 
  url = {https://r-nimble.org} 
}

@article{neal2000markov,
  title={Markov chain sampling methods for {D}irichlet process mixture models},
  author={Neal, Radford M.},
  journal={Journal of Computational and Graphical Statistics},
  volume={9},
  number={2},
  pages={249--265},
  year={2000},
  publisher={Taylor \& Francis}
}

@article{krnjajic2008,
title = {Parametric and nonparametric Bayesian model specification: A case study involving models for count data},
author = {Krnjaji{\'c}, Milovan and Kottas, Athanasios and Draper, David},
journal = {Computational Statistics \& Data Analysis},
volume = {52},
number = {4},
pages = {2110-2128},
year = {2008},
publisher = {Elsevier},
issn = {0167-9473},
doi = {https://doi.org/10.1016/j.csda.2007.07.010},
url = {https://www.sciencedirect.com/science/article/pii/S016794730700268X},
}

@article{laud1995,
    author = {Laud, Purushottam W. and Ibrahim, Joseph G.},
    title = {Predictive Model Selection},
    journal = {Journal of the Royal Statistical Society: Series B (Methodological)},
    volume = {57},
    number = {1},
    pages = {247-262},
    year = {1995},
    month = {12},
    issn = {0035-9246},
    doi = {10.1111/j.2517-6161.1995.tb02028.x},
    url = {https://doi.org/10.1111/j.2517-6161.1995.tb02028.x}
}

@article{xia2012continental,
  title={Continental-scale water and energy flux analysis and validation for the North American Land Data Assimilation System project phase 2 (NLDAS-2): 1. Intercomparison and application of model products},
  author={Xia, Youlong and Mitchell, Kenneth and Ek, Michael and Sheffield, Justin and Cosgrove, Brian and Wood, Eric and Luo, Lifeng and Alonge, Charles and Wei, Helin and Meng, Jesse and others},
  journal={Journal of Geophysical Research: Atmospheres},
  volume={117},
  number={D3},
  year={2012},
  publisher={Wiley Online Library}
}

@article{mitchell2004multi,
  title={The multi-institution North American Land Data Assimilation System (NLDAS): Utilizing multiple GCIP products and partners in a continental distributed hydrological modeling system},
  author={Mitchell, Kenneth E and Lohmann, Dag and Houser, Paul R and Wood, Eric F and Schaake, John C and Robock, Alan and Cosgrove, Brian A and Sheffield, Justin and Duan, Qingyun and Luo, Lifeng and others},
  journal={Journal of Geophysical Research: Atmospheres},
  volume={109},
  number={D7},
  year={2004},
  publisher={Wiley Online Library}
}

@article{feng2014composite,
  title={Composite likelihood estimation for models of spatial ordinal data and spatial proportional data with zero/one values},
  author={Feng, Xiaoping and Zhu, Jun and Lin, Pei-Sheng and Steen-Adams, Michelle M},
  journal={Environmetrics},
  volume={25},
  number={8},
  pages={571--583},
  year={2014},
  publisher={Wiley Online Library}
}

@article{higgs2010clipped,
  title={A clipped latent variable model for spatially correlated ordered categorical data},
  author={Higgs, Megan Dailey and Hoeting, Jennifer A},
  journal={Computational Statistics \& Data Analysis},
  volume={54},
  number={8},
  pages={1999--2011},
  year={2010},
  publisher={Elsevier}
}

@article{erkanli1993,
  title={A Bayesian analysis of ordinal data using mixtures},
  author={Erkanli, Alaattin and Stangl, Dalene and M{\"u}ller, Peter},
  journal={ASA Proceedings of the Section on Bayesian Statistical Science},
  pages={51--56},
  year={1993},
  publisher={American Statistical Association}
}

@article{kottas2005,
  title={Nonparametric Bayesian Modeling for Multivariate Ordinal Data},
  author={Kottas, Athanasios and M{\"u}ller, Peter and Quintana, Fernando A},
  journal={Journal of Computational and Graphical Statistics},
  volume={14},
  number={3},
  pages={610--625},
  year={2005},
  publisher={Taylor \& Francis},
  doi = {10.1198/106186005X63185},
URL = {https://doi.org/10.1198/106186005X63185},
eprint = {https://doi.org/10.1198/106186005X63185}  
}

@article{deYoreo2018,
  title={Bayesian Nonparametric Modeling for Multivariate Ordinal Regression},
  author={DeYoreo, Maria and Kottas, Athanasios},
  journal={Journal of Computational and Graphical Statistics},
  volume={27},
  number={1},
  pages={71--84},
  year={2018},
  publisher={Taylor \& Francis},
  doi = {10.1080/10618600.2017.1316280},
  URL = {https://doi.org/10.1080/10618600.2017.1316280},
eprint = {https://doi.org/10.1080/10618600.2017.1316280}
}

@article{diLucca2013,
author={Di Lucca, Maria Anna and Guglielmi, Alessandra and M{\"u}ller, Peter and Quintana, Fernando A},
title = {{A Simple Class of Bayesian Nonparametric Autoregression Models}},
volume = {8},
journal = {Bayesian Analysis},
number = {1},
publisher = {International Society for Bayesian Analysis},
pages = {63 -- 88},
year = {2013},
doi = {10.1214/13-BA803},
URL = {https://doi.org/10.1214/13-BA803}
}

@article{nieto-barajas2014,
author={Nieto-Barajas, Luis E and Contreras-Crist{\'a}n, Alberto},
title = {{A Bayesian Nonparametric Approach for Time Series Clustering}},
volume = {9},
journal = {Bayesian Analysis},
number = {1},
publisher = {International Society for Bayesian Analysis},
pages = {147 -- 170},
year = {2014},
doi = {10.1214/13-BA852},
URL = {https://doi.org/10.1214/13-BA852}
}

@article{mozdzen2022,
author={Mozdzen, Alexander and Cremaschi, Andrea and Cadonna, Annalisa and Guglielmi, Alessandra and Kastner, Gregor},
title = {Bayesian modeling and clustering for spatio-temporal areal data: An application to Italian unemployment},
journal = {Spatial Statistics},
volume = {52},
pages = {100715},
year = {2022},
issn = {2211-6753},
doi = {https://doi.org/10.1016/j.spasta.2022.100715},
url = {https://www.sciencedirect.com/science/article/pii/S2211675322000768},
publisher={Elsevier}
}

@article{grazian2024,
author={Grazian, Clara},
title = {{Spatio-Temporal Stick-Breaking Process}},
volume = {20},
journal = {Bayesian Analysis},
number = {3},
publisher = {International Society for Bayesian Analysis},
pages = {763 -- 794},
year = {2025},
doi = {10.1214/24-BA1419},
URL = {https://doi.org/10.1214/24-BA1419}
}

@article{aiello2025,
title={Bayesian nonparametric clustering for spatio-temporal data, with an application to air pollution},
author={Aiello, Luca and Argiento, Raffaele and Legramanti, Sirio and Paci, Lucia},
journal={arXiv preprint arXiv:2505.24694},
year={2025},
eprint={2505.24694},
archivePrefix={arXiv},
primaryClass={stat.ME},
url={https://arxiv.org/abs/2505.24694}
}
